\documentclass[12pt, a4paper]{article}
\usepackage[dvipdfmx]{graphicx}
\usepackage{amssymb}
\usepackage{amsmath}
\usepackage{bm}
\usepackage{color}
\usepackage{cite}
\usepackage{slashed}
\usepackage{subfigure}
\usepackage{epstopdf}            
\usepackage{epsfig}
\usepackage{here}
            
\renewcommand{\thefootnote}{\fnsymbol{footnote}}
\numberwithin{equation}{section} 
\def\beq#1\eeq{\begin{align}#1\end{align}}

\RequirePackage{xspace}
\def\Bbar    {\kern 0.18em\overline{\kern -0.18em B}{}\xspace}
\def\Bb      {\ensuremath{\Bbar}\xspace}

\usepackage{caption}
\definecolor{BlueViolet}{rgb}{0.2, 0.00, 0.7}
\definecolor{Blue}{rgb}{0.15, 0.00, 0.9}
\usepackage[
colorlinks=true, linkcolor=Blue,citecolor=Blue,urlcolor=BlueViolet]{hyperref}

\begin{document}

\begin{titlepage}

\begin{center}

\hfill TTP18--037,
 ZU-TH-39/18\\
\hfill November 2018


{\LARGE\bf $\boldsymbol{D^{\ast}}$ polarization vs. $\boldsymbol{R_{D^{(\ast)}}}$ anomalies \\ \vspace{.1 in}
in the leptoquark models
}

\vskip .2in

{\large 
Syuhei Iguro$^{\rm (a)}$,
Teppei Kitahara$^{\rm (b,c,d,e,f)}$,
Yuji Omura$^{\rm (c)}$, \\ 
\vspace{.2cm}}
{\large 
Ryoutaro Watanabe$^{\rm (g)}$, 
 and
  Kei Yamamoto$^{\rm (h,i)}$
}

\vskip 0.25in

$^{\rm (a)}${\it 
Department of Physics,
Nagoya University, Nagoya 464-8602, Japan}

\vskip 0.1in

$^{\rm (b)}${\it 
Institute for Advanced Research, Nagoya University,
Furo-cho Chikusa-ku, Nagoya, Aichi, 464-8602 Japan}

\vskip 0.1in

$^{\rm (c)}${\it 
Kobayashi-Maskawa Institute for the Origin of Particles and the
Universe, \\ Nagoya University, Nagoya 464-8602, Japan}

\vskip 0.1in

$^{\rm (d)}${\it 
Institute for Theoretical Particle Physics (TTP), Karlsruhe Institute of Technology, Engesserstra{\ss}e 7, D-76128 Karlsruhe, Germany}
 
\vskip 0.1in
 
$^{\rm (e)}${\it
Institute for Nuclear Physics (IKP), Karlsruhe Institute of
Technology, Hermann-von-Helmholtz-Platz 1, D-76344
Eggenstein-Leopoldshafen, Germany} 

\vskip 0.1in

$^{\rm (f)}${\it 
Physics Department, Technion--Israel Institute of Technology, Haifa 3200003, Israel
}

\vskip 0.1in

$^{\rm (g)}${\it 
INFN, Sezione di Roma Tre, 00146 Rome, Italy
}
\vskip 0.1in

$^{\rm (h)}${\it 
Physik-Institut, Universit\"at  Z\"urich, CH-8057 Z\"urich, Switzerland
}
\vskip 0.1in

$^{\rm (i)}${\it 
Graduate School of Science, Hiroshima University, Higashi-Hiroshima 739-8526, Japan
}

\end{center}

\begin{abstract}
Polarization measurements in $\Bb \to D^{(\ast)} \tau \overline{\nu}$ are useful to check consistency in new physics explanations for the $R_{D}$ and $R_{D^{\ast}}$ anomalies.
In this paper, we investigate the $D^{\ast}$ and $\tau$ polarizations 
and focus on the new physics contributions to the fraction of a longitudinal $D^{\ast}$ polarization ($F_{L}^{D^{\ast}}$), 
which is recently measured by the Belle collaboration $F_{L}^{D^{\ast}} = 0.60 \pm 0.09$, 
in model-independent manner and in each single leptoquark model (${\rm R}_2$, ${\rm S}_1$ and ${\rm U}_1$) that can naturally explain the  $R_{D^{(\ast)}}$ anomalies.
It is found that $\mathcal{B}(B_c^{+} \to \tau^{+} \nu)$ severely restricts deviation from the Standard Model (SM) prediction of $F_{L,\,\textrm{SM}}^{D^{\ast}} = 0.46 \pm 0.04$ in the leptoquark models:
$[0.43, 0.44]$, $[0.42, 0.48]$, and $[0.43, 0.47]$ are predicted as a range of $F_{L}^{D^{\ast}}$ for the ${\rm R}_2$, ${\rm S}_1$, and ${\rm U}_1$ leptoquark models, respectively, 
where the current data of $R_{D^{(\ast)}}$ is satisfied at $1\,\sigma$ level. 
It is also shown that the $\tau$ polarization observables can much deviate from the SM predictions.
The Belle II experiment, therefore, can check such correlations between $R_{D^{(\ast)}}$ and the polarization observables, and discriminate among the leptoquark models. 
\end{abstract}
\end{titlepage}

\setcounter{page}{1}
\renewcommand{\thefootnote}{\#\arabic{footnote}}
\setcounter{footnote}{0}

\section{Introduction}
\label{sec:intro}

Semi-leptonic $B$ meson decays have been investigated to test the Standard Model (SM) since the CLEO, BaBar, and Belle experiments were established. 
In the processes, the SM predicts the specific flavor structure: 
the quark mixing is suppressed by the Cabbibo-Kobayashi-Maskawa (CKM) matrix elements~\cite{Cabibbo:1963yz,Kobayashi:1973fv} and the dependence on the lepton flavor in the final state is universal in the predictions. 
Therefore, steady efforts have been made to measure them with high accuracy.
The measurements are significant for not only the test of the SM but also probing New Physics (NP).

On recent years, the semi-tauonic processes, $\Bb \to D^{(\ast)} \tau \overline{\nu}$, have come under the spotlight  
since the BaBar~\cite{Lees2012xj,Lees2013uzd}, Belle~\cite{Huschle2015rga,Sato2016svk,Hirose2016wfn} and LHCb~\cite{Aaij2015yra,Aaij:2017uff} experiments 
have shown discrepancies between their data 
and the SM predictions in the measurements of 
 \begin{align}
  R_D = \frac{\mathcal{B}(\Bb\rightarrow D \tau \bar{\nu})}{\mathcal{B}(\Bb\rightarrow D \ell\bar{\nu})} \,, ~~~~
  R_{D^{*}} = \frac{\mathcal{B}(\Bb\rightarrow D^{*} \tau \bar{\nu})}{\mathcal{B}(\Bb\rightarrow D^{*} \ell\bar{\nu})} \,,   
\end{align}
where $\ell =e,\mu$.

The current situation on the experimental values and the SM predictions are summarized in Ref.~\cite{HFLAV} 
as $R_D^\textrm{exp}=0.407 \pm 0.046$, $R_{D^{\ast}}^\textrm{exp}=0.306 \pm 0.015$, $R_D^\textrm{SM}=0.299 \pm 0.003$, and $R_{D^{\ast}}^\textrm{SM}=0.258 \pm 0.005$. 
Hence the combined deviation is now $3.8\,\sigma$, referred to as $R_{D^{(*)}}$ anomalies. 
The surprising fact is that these decay processes are described by the tree-level amplitude in the SM and thus such a large discrepancy implies large unknown effects in the processes.

Motivated by those results, a lot of studies have been done from different points of view:
re-evaluations of the form factors in the SM predictions, studies to accommodate the $R_{D^{(*)}}$ anomalies in NP models, and utilities of other observables than $R_{D^{(*)}}$ to probe NP effects. 
An overview of the above points, based on recent developments, can be summarized as follows: 
\begin{itemize}
 \item 
 For the SM predictions, the heavy quark effective theory (HQET) has been applied to the form factors of the $\Bb\rightarrow D^{(*)}$ transitions~\cite{Caprini:1997mu}. 
 In Refs.~\cite{Bernlochner:2017jka,Jaiswal:2017rve}, $\mathcal O(\Lambda_\text{QCD}/M_Q)$ and $\mathcal O(\alpha_s)$ corrections to the form factors in HQET are obtained. 
 In Refs.~\cite{Bigi:2016mdz,Bigi:2017jbd}, another approach has been considered by taking the Boyd-Grinstein-Lebed parameterization~\cite{Boyd:1997kz}. 
 Both of the two approaches enable us to evaluate the SM values with 1\% level of the uncertainties as shown above. 
 In Ref.~\cite{deBoer:2018ipi}, corrections from soft-photon effects are calculated and then one finds that it gives up to $3 \,\text{--}\, 4\%$ amplification of $R_D^{\textrm{SM}}$. 
 \item
 The NP studies are summarized below:
 \begin{itemize}
 \item
 One possible NP candidate to explain this anomaly was a charged scalar boson~\cite{
 Crivellin2012ye,Celis:2012dk,Tanaka2012nw,Ko:2012sv,Crivellin:2013wna,Crivellin:2015hha,Kim:2015zla,Cline:2015lqp,Celis:2016azn,
 Ko:2017lzd,Iguro:2017ysu,Fuyuto:2017sys,Iguro:2018qzf,Iguro:2018oou,Martinez:2018ynq,Fraser:2018aqj,Li:2018rax
 }. 
 Regardless of the detail of the model, however, it has been turned out that this kind of scenario~(\textrm{i.e.}, with scalar mediator) becomes inconsistent with the bound from the $B_c^+$ lifetime~\cite{Beneke:1996xe, Li:2016vvp, Alonso:2016oyd, Celis:2016azn, Akeroyd:2017mhr}. 
 It is also found that the direct search for $\tau \nu$ resonance at the LHC gives a bound which could be more stringent depending on the mass and the branching ratio of the charged scalar boson \cite{Iguro:2018fni}. 
 \item 
 A charged vector boson ($W'$) could be in this game~\cite{He:2012zp,Greljo:2015mma,Boucenna:2016wpr,He:2017bft,Cvetic:2017gkt,Asadi:2018wea,Greljo:2018ogz,Babu:2018vrl}. 
 In order to introduce such a new vector field, we need additional gauge symmetry, which also leads to an additional neutral vector boson ($Z'$) in general. 
 With this additional ingredient, one can discuss a correlation between $\Bb\rightarrow D^{(*)} \tau \bar{\nu}$ and other processes such as $B \to K^{(*)}\mu^+\mu^-$.\footnote
 {
 A simultaneous explanation of the anomalies in $b \to c \tau\nu$ ($R_{D^{(*)}}$) and 
 $b \to s \mu\mu$ ($R_{K^{(*)}}$~\cite{Aaij:2014ora,Aaij:2017vbb}, 
 $B \to K^*\mu^+\mu^-$~\cite{Aaij:2013qta,Aaij:2015oid,Abdesselam:2016llu,ATLAS:2017dlm,CMS:2017ivg}, 
 $B_s^0\to\phi\mu^+\mu^-$~\cite{Aaij:2013aln,Aaij:2015esa}) is another direction for the NP study 
 (see Ref.~\cite{Kumar:2018kmr} for example), which will not be the subject in this paper.  
 }
 \item 
 A scalar or vector boson that couples to a quark and lepton pair~\cite{Buchmuller:1986zs}, namely leptoquark (LQ), is another candidate as will be discussed in detail later. 
 It has been already pointed out that three types of LQ models can accommodate the $R_{D^{(*)}}$ anomalies~\cite{Sakaki:2013bfa}. 
 \end{itemize}
\item
Other observables of $\Bb\rightarrow D^{(*)} \tau \bar{\nu}$ have been examined in order to probe/distinguish NP effects/scenarios. 
At the coming Belle~II experiment, a large amount of signal events will be available and thus distributions of the processes would be useful for this purpose. 
In Refs.\cite{Sakaki:2014sea,Kou:2018nap}, it is pointed out that $5\,\text{ab}^{-1}$ data of the $q^2 = (p_B-p_{D^{(*)}})^2$ distribution expected at the Belle~II can distinguish some NP scenarios that can explain the present $R_{D^{(*)}}$ anomalies. 
The polarizations of $\tau$ and $D^*$ are also good candidates to test the NP scenarios~\cite{Tanaka:2010se,Tanaka:2012nw}. 
They reflect the spin structure of the interaction in $\Bb\rightarrow D^{(*)} \tau \bar{\nu}$, and could be affected by NP, 
{\it e.g.}, see Refs.\cite{Tanaka:2010se,Tanaka:2012nw,Sakaki:2013bfa,Alonso:2016gym,Alok:2016qyh,Ivanov:2017mrj,Alonso:2017ktd}. 
Relations between the $R_{D^{(*)}}$ anomalies and $|V_{cb}|$ determination with a tensor operator have also been discussed~\cite{ Biancofiore:2013ki, Colangelo:2016ymy, Colangelo:2018cnj}.
\end{itemize}

In this paper, we focus on the $D^*$ polarization following the new observation from the Belle experiment in which their first preliminary result of the longitudinal polarization $F_L^{D^*}$ has been given as~\cite{Adamczyk} 
\beq
F_{L}^{D^{\ast}} = 0.60 \pm 0.08 ({\rm stat.}) \pm 0.035 ({\rm syst.}).
\label{eq:FLDst_exp}
\eeq
This is then compared with the SM prediction: $F_{L,\,\textrm{SM}}^{D^{\ast}} = 0.46 \pm 0.04$~\cite{Alok:2016qyh}. 
Although they are consistent at $1.5\,\sigma$, the point here is that the experimental value is larger than the SM one. 
Indeed, this is an opposite correlation with the present 
$R_{D^{*}}$ 
anomaly in 
the 
presence of one NP effective operator for $b \to c \tau \nu$ as shown in Ref.~\cite{Tanaka:2012nw} 
except for scalar NP scenarios.

In the light of this situation, we investigate relations among $R_{D}$, $R_{D^{\ast}}$, and $F_{L}^{D^{\ast}}$ in the LQ models 
that induce more than two effective operators, and see if they could accommodate the present data.  
We will begin with obtaining numerical formulae in terms of Wilson coefficients for NP operators by taking into account the recent development on the form factors.
Then, we will show possible reaches of $F_{L}^{D^{\ast}}$ when we take into account the $R_{D^{(*)}}$ anomalies in the LQ models. 
We will also point out that the $\tau$ polarizations are useful to distinguish the LQ models based on sensitivities expected at the Belle II experiment.

This paper is organized as follows. 
In Sec.~\ref{sec2}, we put the numerical formulae for the relevant observables in terms of the effective Hamiltonian.
We also summarize the case for single operator analysis. 
In Sec.~\ref{sec3}, based on the generic study with renormalization-group running effects, 
we obtain relations among $R_{D}$, $R_{D^{\ast}}$, and $F_{L}^{D^{\ast}}$ in the LQ models and discuss their potential to explain the present data. 
Relations to the $\tau$ polarizations are also discussed.
Finally, we conclude our study in Sec.~\ref{conclusion}.

\section{Formulae for the observables}
\label{sec2}

At first, we describe general NP contributions in terms of the effective Hamiltonian.
The operators relevant to $\Bb \to D^{(\ast)} \tau \overline{\nu}$ are described as\footnote{
Another convention used in the literature \cite{Feruglio:2018fxo, Angelescu:2018tyl} is related as
$C_{V_1} = g^{\tau}_{V_L}$, $C_{V_2}= g^{\tau}_{V_R}$, $C_{S_1} = g^{\tau}_{S_R}$, $C_{S_2}= g^{\tau}_{S_L}$, and $C_{T}=g^{\tau}_{T}$. 
} 
\begin{align}
\label{eq:Hamiltonian}
{\cal {H}}_{\rm{eff}}=\frac{4G_F}{\sqrt2}V_{cb}\biggl[ (1+C_{V_1})O_{V_1}+C_{V_2}O_{V_2}+C_{S_1}O_{S_1}+C_{S_2}O_{S_2}+C_{T}O_{T}\biggl],
\end{align}
at the scale $\mu = \mu_b = 4.2\,\text{GeV}$ with
\begin{align}
O_{V_1}&=(\overline{c} \gamma^\mu P_Lb)(\overline{\tau} \gamma_\mu P_L \nu_{\tau}), ~~~
O_{V_2}=(\overline{c} \gamma^\mu P_Rb)(\overline{\tau} \gamma_\mu P_L \nu_{\tau}),\nonumber \\
O_{S_1}&=(\overline{c}  P_Rb)(\overline{\tau} P_L \nu_{\tau}),~~~~~~~~~\,
O_{S_2}=(\overline{c}  P_Lb)(\overline{\tau} P_L \nu_{\tau}), \nonumber \\
O_{T}&=(\overline{c}  \sigma^{\mu\nu}P_Lb)(\overline{\tau} \sigma_{\mu\nu} P_L \nu_{\tau})\label{eq:operator},
\end{align}
where $P_L=(1-\gamma_5)/2$ and $P_R=(1+\gamma_5)/2$. 
Note that the SM prediction is given by $C_{X} = 0$ for $X=V_{1,2}$, $S_{1,2}$, and $T$ in this description. 
We assume that the neutrino is always left-handed and third-generation ($\nu_\tau$). 
In LQ models, the neutrino flavor could be first- or second-generation ($\nu_{\mu,e}$) as seen in the next section. 
In principle,  one can translate $C_X$ into that of $\nu_{\mu,\,e}$.
Possibilities of the light  sterile neutrinos are discussed in Refs.~\cite{Iguro:2018qzf, He:2017bft, Asadi:2018wea, Greljo:2018ogz, Babu:2018vrl,Robinson:2018gza,Azatov:2018kzb}.

In this work, we follow analytic forms for the decay rates obtained in Refs.~\cite{Sakaki:2013bfa,Sakaki:2014sea}. 
As for all the form factors in both SM and NP amplitudes, we universally adopt the recent development taken in Ref.~\cite{Bernlochner:2017jka} such that a proper manner of the HQET expansion can be evaluated. 
To be precise, we have adopted the fit scenario ``$\textrm{L}_{w\geq 1}+$SR" \cite{Bernlochner:2017jka}, 
where the HQET expansion is evaluated at the matching scale $\mu = \sqrt{m_b m_c}$ with QCD.
According to Eq.~(A5) of Ref.~\cite{Bernlochner:2017jka}, we have evaluated observables at the scale $\mu = \mu_b$, as defined in the effective Hamiltonian of Eq.~\eqref{eq:Hamiltonian}.
In the end, we find the following numerical formulae 
\begin{align}
 \label{eq:RD}
 \frac{R_D}{R_{D}^\textrm{SM}} =
 & ~|1+C_{V_1}+C_{V_2}|^2  + 1.02|C_{S_1}+C_{S_2}|^2 + 0.90|C_{T}|^2 \nonumber \\
 & + 1.49\textrm{Re}[(1+C_{V_1}+C_{V_2})(C_{S_1}^*+C_{S_2}^*)]  + 1.14\textrm{Re}[(1+C_{V_1}+C_{V_2})C_{T}^*] \,, \\[1.0em]
 \label{eq:RDs}
 \frac{ R_{D^{\ast}}}{R_{D^\ast}^\textrm{SM}} =
 & ~|1+C_{V_1}|^2 + |C_{V_2}|^2  + 0.04|C_{S_1}-C_{S_2}|^2 + 16.07|C_{T}|^2 \nonumber \\
 & -1.81\textrm{Re}[(1+C_{V_1})C_{V_2}^*]  + 0.11\textrm{Re}[(1+C_{V_1}-C_{V_2})(C_{S_1}^*-C_{S_2}^*)] \nonumber \\[0.5em] 
 & -5.12\textrm{Re}[(1+C_{V_1})C_{T}^*] + 6.66\textrm{Re}[C_{V_2}C_{T}^*] \,, 
\end{align}
which can be compared with those in the recent literature~\cite{Feruglio:2018fxo, Asadi:2018wea,Blanke:2018yud}.\footnote{Differences of the numerical results stem from an input and method to describe the form factors. } 
Using our code, we obtained the SM predictions as $R_{D}^\textrm{SM} = 0.300$ and $R_{D^\ast}^\textrm{SM} = 0.256$, which are well consistent with Ref.~\cite{Bernlochner:2017jka}.

Note that our  values of $R_{D^{(\ast)}}$ and the following polarization observables are valid up to $\mathcal{O}(\Lambda_{\rm QCD} /m_{c,b})$ and $\mathcal{O}(\alpha_s)$ within uncertainties\footnote{
Recently, Ref.~\cite{Jung:2018lfu} has suggested that a higher order contribution of $\mathcal O(\Lambda_{\rm QCD}^2 /m_{c,b}^2)$ may have an impact on the evaluation. 
} from the input parameters~\cite{Bernlochner:2017jka}. 
We also emphasize that we have taken care of the scale for the Wilson coefficients and that for the HQET expansion to be $\mu = \mu_b = 4.2\,\text{GeV}$.
Although the SM operator is independent of such a scale, the NP operators do depend on it. 
For example, the coefficient of the $|C_T|^2$ term in $R_{D^\ast}/R_{D^\ast}^\textrm{SM}$ is found to be $17.24$ at the scale $\mu = \sqrt{m_b m_c}=2.6\,\text{GeV}$, 
whereas $16.07$ at $\mu = \mu_b = 4.2\,\text{GeV}$ as shown in our result. 
This difference is indeed compensated with the running effect on the Wilson coefficient given as $C_T(\mu=2.6\,\text{GeV}) = 0.97\,C_T(\mu=4.2\,\text{GeV})$.

In a similar way, we can also calculate the polarizations in $\Bb \to D^{(\ast)} \tau \overline{\nu}$. 
The $D^{\ast}$ polarization is defined as the fraction of a longitudinal mode for the $D^{\ast}$ meson, namely, 
\beq
F_{L}^{D^{\ast}} &= \frac{\Gamma(\Bb \to D^{\ast}_L \tau \overline{\nu} )}{\Gamma(\Bb \to D^{\ast} \tau \overline{\nu} )}
 = \frac{\Gamma(\Bb \to D^{\ast}_L \tau \overline{\nu} )}{\Gamma(\Bb \to D^{\ast}_{L} \tau \overline{\nu} ) + \Gamma(\Bb \to D^{\ast}_{T} \tau \overline{\nu} )}, 
\eeq
where $D^{\ast}_{L(T)}$ denotes the longitudinal (transverse) mode of the $D^{\ast}$ meson. 
For the numerical formula, we obtain  
\begin{align}
{F_L^{D^*} \over F_{L,\,\textrm{SM}}^{D^{\ast}}} = 
 & \left({R_{D^*} \over R_{D^*}^\text{SM}}\right)^{-1} \!\!\!\times \Big( |1+C_{V_1}-C_{V_2}|^2  + 0.08|C_{S_1}-C_{S_2}|^2 + 7.02|C_{T}|^2 \nonumber \\
 & + 0.24\textrm{Re}[(1+C_{V_1}-C_{V_2})(C_{S_1}^*-C_{S_2}^*)]  -4.37\textrm{Re}[(1+C_{V_1}-C_{V_2})C_{T}^*] \Big) . 
\end{align}
Here the SM prediction is $F_{L,\,\textrm{SM}}^{D^{\ast}} = 0.453$, which is consistent with Ref.~\cite{Alok:2016qyh}.

For the $\tau$ polarization asymmetries along the longitudinal directions of the $\tau$ leptons in  $\Bb\rightarrow D \tau \bar{\nu}$   and $\Bb\rightarrow D^{\ast} \tau \bar{\nu}$, 
we obtain
\begin{align}
{P_\tau^{D} \over P_{\tau,\,\textrm{SM}}^{D}} = 
 & \left({R_D \over R_D^\text{SM}}\right)^{-1} \!\!\!\times \Big( |1+C_{V_1}+C_{V_2}|^2  + 3.18|C_{S_1}+C_{S_2}|^2 + 0.18|C_{T}|^2 \nonumber \\
 & + 4.65\textrm{Re}[(1+C_{V_1}+C_{V_2})(C_{S_1}^*+C_{S_2}^*)]  -1.18\textrm{Re}[(1+C_{V_1}+C_{V_2})C_{T}^*] \Big)  \,, 
\end{align}
and
 \begin{align}
{P_\tau^{D^{\ast}} \over P_{\tau,\,\textrm{SM}}^{D^{\ast}}} = 
 & \left({R_{D^{\ast}} \over R_{D^{\ast}}^\text{SM}}\right)^{-1} \!\!\!\times \Big(
 |1+C_{V_1}|^2  + |C_{V_2}|^2  - 0.07|C_{S_1}- C_{S_2}|^2 - 1.86 |C_{T}|^2 \nonumber \\
 & 
 - 1.77 \textrm{Re}[(1+C_{V_1})C_{V_2}^{\ast}]  
 - 0.22 \textrm{Re}[(1+C_{V_1}- C_{V_2})(C_{S_1}^* -C_{S_2}^*)]  \nonumber \\[0.5em]
 & 
 - 3.37 \textrm{Re}[(1+C_{V_1}) C_{T}^*]
 + 4.37  \textrm{Re}[C_{V_2} C_{T}^*] \Big), 
 \end{align} 
respectively. 
The definitions of $P_\tau^{D}$ and $P_\tau^{D^{\ast}}$ are given in Refs.~\cite{Tanaka:2012nw,Asadi:2018sym}.
Based on the present framework for the form factors, we obtain the SM predictions as $P_{\tau,\,\textrm{SM}}^{D} = 0.320$ and $P_{\tau,\,\textrm{SM}}^{D^{\ast}} = -0.507$. 
Note that the polarizations are measurable by analyzing angular and/or energy distributions, {\it e.g.}, see Refs.~\cite{Tanaka:2010se,Alok:2016qyh}. 
For comparison, $P_{\tau,\,\textrm{SM}}^{D} = 0.325 \pm 0.009$ and $P_{\tau,\,\textrm{SM}}^{D^{\ast}} = - 0.497 \pm  0.013$ are obtained in Ref.~\cite{Hirose2016wfn} (Belle estimation), 
while $P_{\tau,\,\textrm{SM}}^{D} = 0.34 \pm 0.03$ in Ref.~\cite{Alonso:2017ktd}.

Another significant observable for our study is the branching ratio of $B_c^+ \to \tau^+ \nu$. 
As shown in Refs.~\cite{Beneke:1996xe,Alonso:2016oyd, Celis:2016azn}, the constraint on the $B_c^+$ lifetime can be translated to that on $\mathcal{B}(B^+_c \to \tau^+ \nu )$ and then one finds that a large scalar NP effect is disfavored. 
We also take this bound into account by using the analytic formula shown in Ref.~\cite{Watanabe:2017mip}:
\beq
\frac{\mathcal{B}(B^+_c\to \tau^+ \nu)}{\mathcal{B}(B^+_c\to\tau^+ \nu_{\tau})_{\textrm{SM}}} & =
\left| 1+C_{V_1}-C_{V_2}+ \frac{m_{B_c}^2}{m_{\tau} \left( \overline{m}_b + \overline{m}_c\right)}\left(C_{S_1}-C_{S_2}\right)   \right|^2 \\
&
 = \left| 1+C_{V_1}-C_{V_2}+ 4.33  \left( C_{S_1}-C_{S_2} \right)\right|^2,
\eeq
where the $\overline{\textrm{MS}}$ quark masses $\overline{m}_{b,c}$ at scale $\mu_b$ are used \cite{Chetyrkin:2000yt}. 
The SM prediction is $0.023$.

\subsection{Case for single NP operator}
\label{sec:operatoranalysis}

Here we review a model-independent study on the correlation between $R_{D^{(\ast)}}$ and $F_L^{D^{\ast}}$ in the presence of a single NP operator in Eq.~\eqref{eq:Hamiltonian} .

\begin{figure}[tp]
  \begin{center}
    \includegraphics[width=0.45\textwidth]{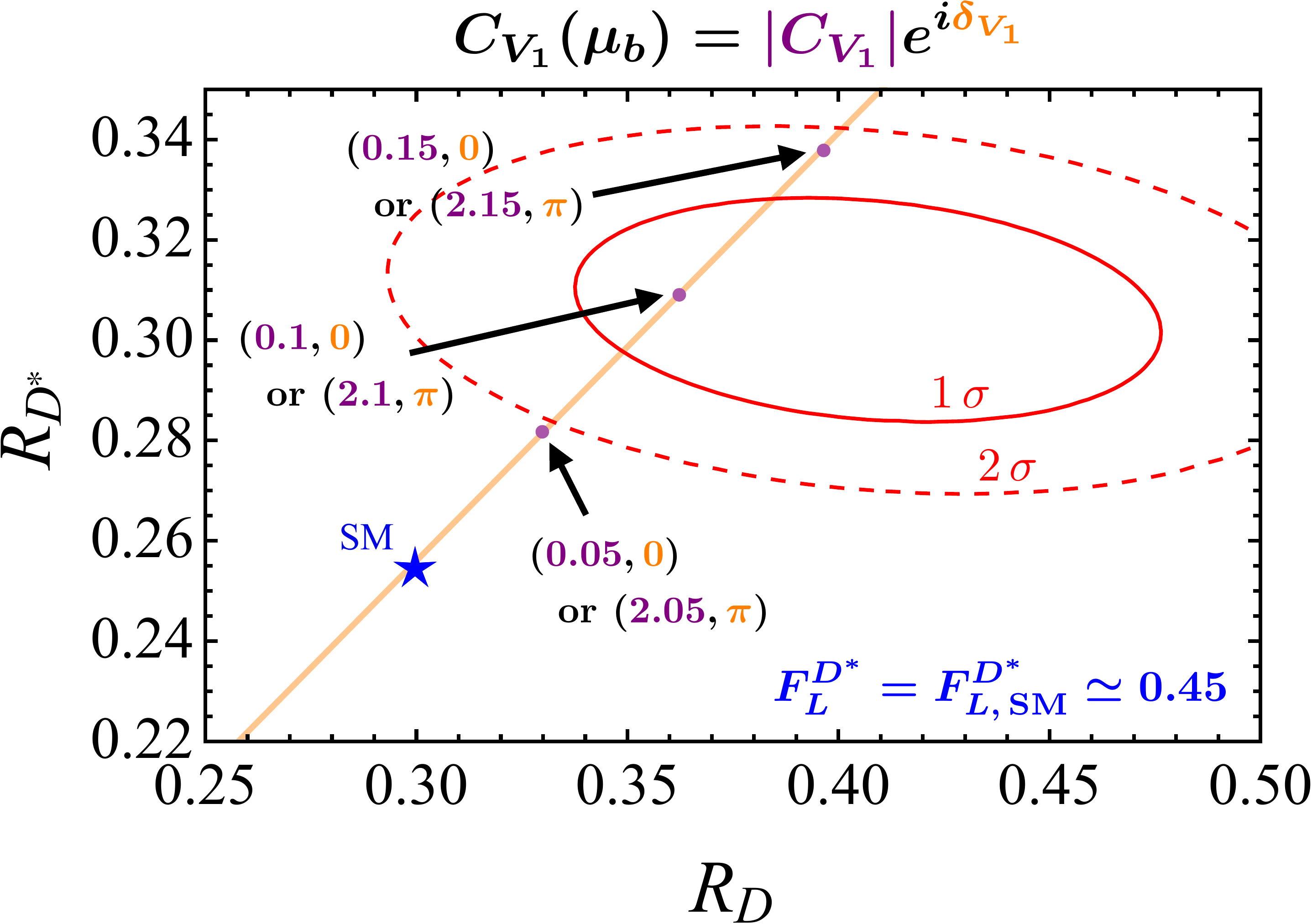}
    \includegraphics[width=0.45\textwidth]{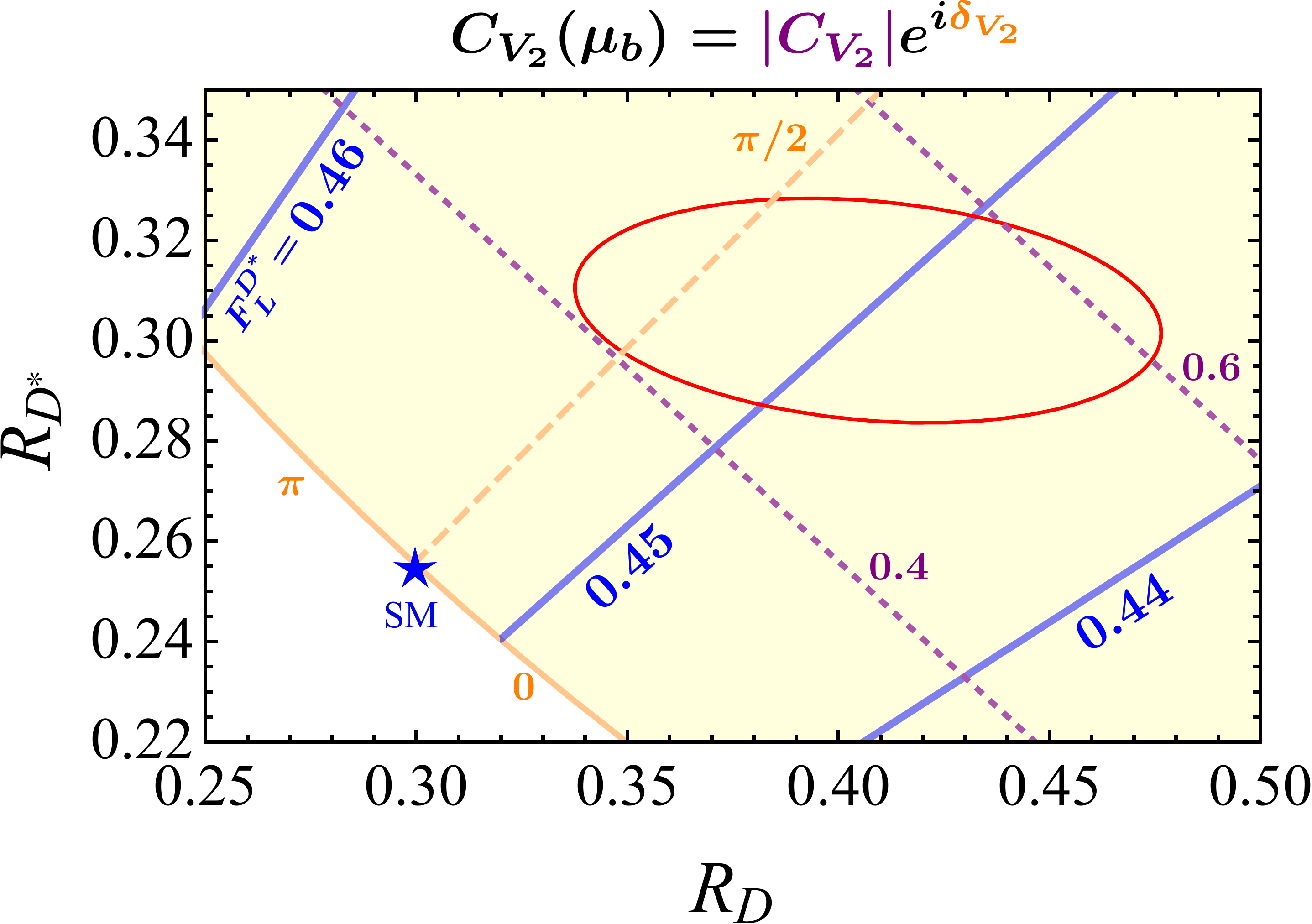}\\ \vspace{.2 in}
        \includegraphics[width=0.45\textwidth]{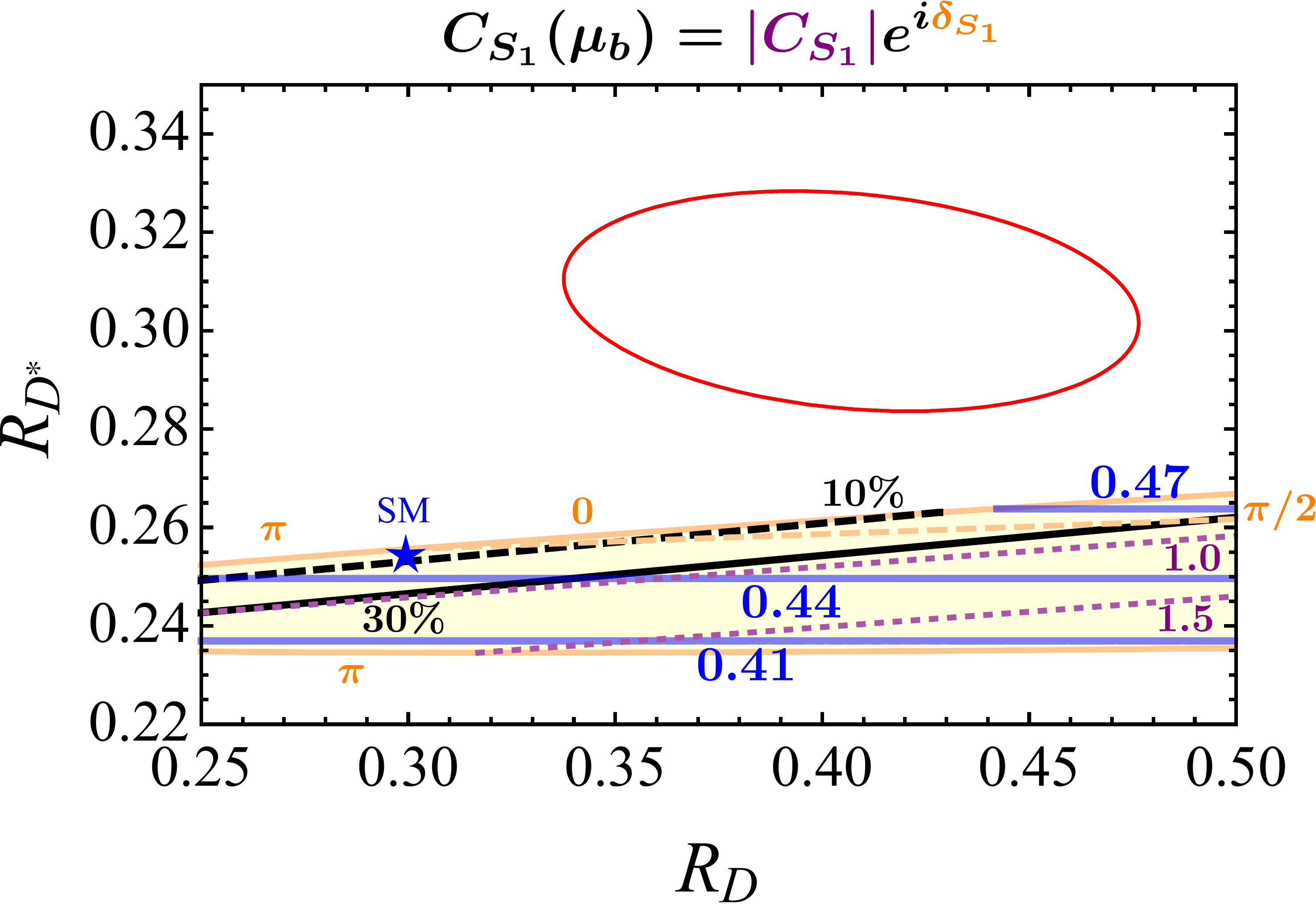}
    \includegraphics[width=0.45\textwidth]{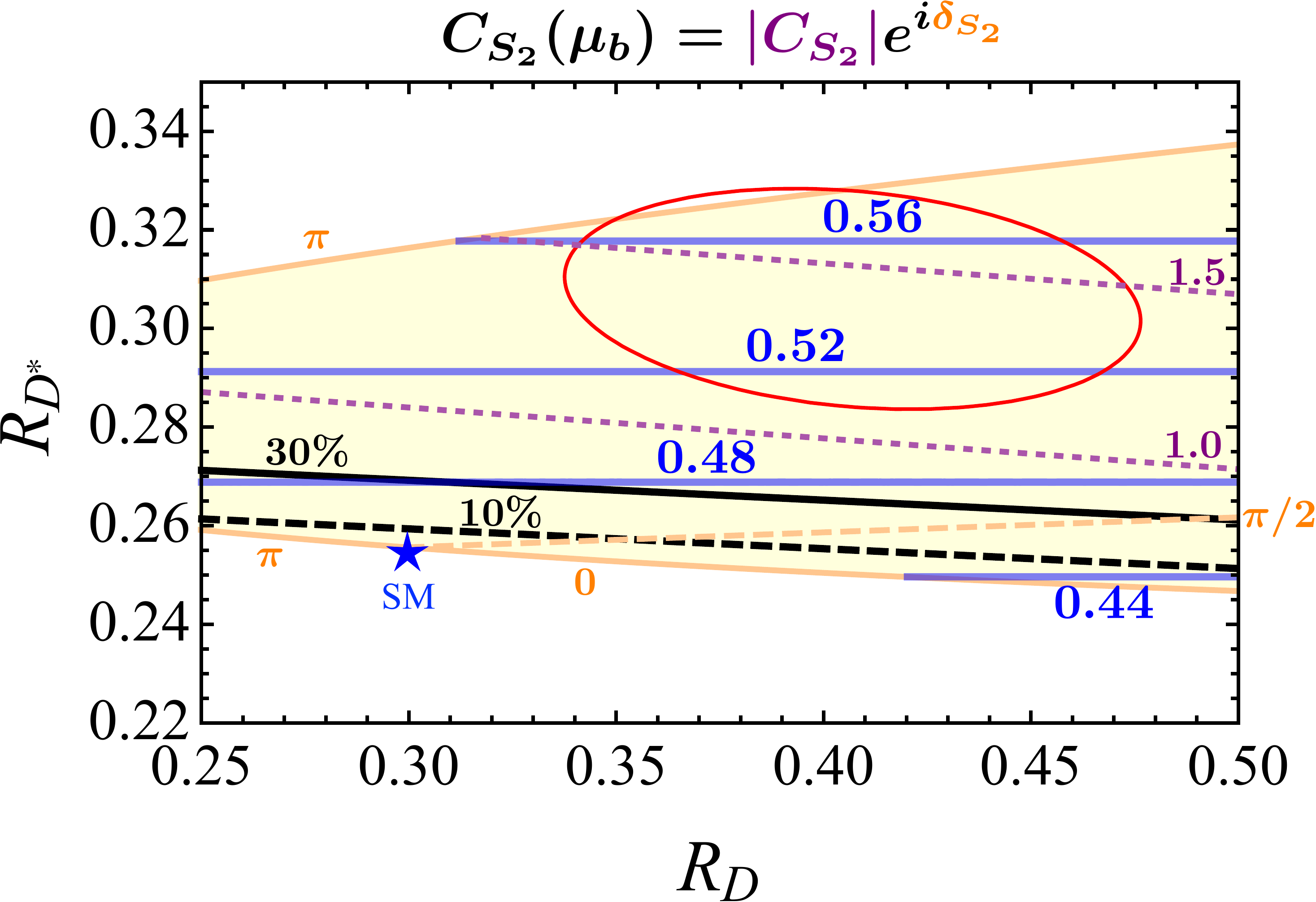}\\ \vspace{.2 in}
        \includegraphics[width=0.45\textwidth]{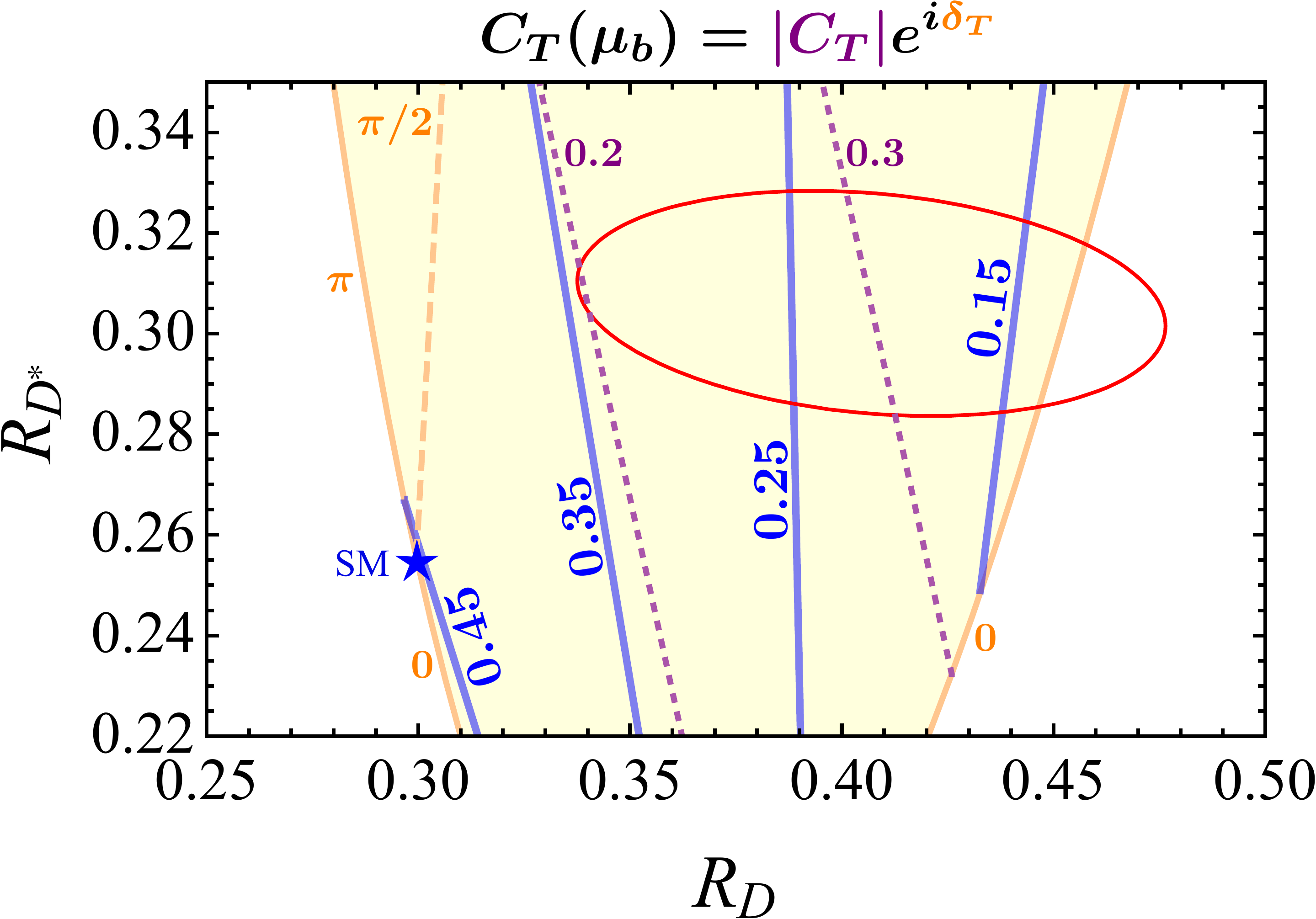}
    \caption{
The contour for $F_L^{D^{\ast}}$ is shown with the blue line on the $R_D$--$R_{D^{\ast}}$ plane in the single NP operator scenario.
The purple and orange lines represent the absolute value and the phase of $C_{X} (\mu_b)$, respectively. 
The constraint from the $B_c^+$ lifetime is drawn by the solid (dashed) black line for $\mathcal{B}(B^+_c \to \tau^+ \nu ) < 0.3$ (0.1).
The world average of the data at $1\,\sigma$ ($2\,\sigma$) is shown by the red (dashed) ellipse.
The SM point is represented by the blue star.
  }
    \label{fig:oneoperator}
  \end{center}
\end{figure}

We parametrize $C_X$ $(X=V_1,V_2,S_1,S_2,T)$ as $C_X=|C_X|e^{i\delta_X}$, and then vary $|C_X|$ and $\delta_X$ (in the range of [0,~$\pi$] for the latter).
As for the $V_1$ case, $|C_{V_1} +1|$ is the only physical parameter and hence we take $C_{V_1}$ to be real for simplicity.
In Fig.~\ref{fig:oneoperator}, the $F_L^{D^{\ast}}$ contour is shown with the blue line on the $R_D$--$R_{D^{\ast}}$ plane for each single operator: $O_{V_1}$, $O_{V_2}$, $O_{S_1}$, $O_{S_2}$, and $O_T$, 
where the shaded region in yellow is achievable with the NP operator and the red (dashed) ellipse stands for the world average of the present data at the $1\,\sigma$ ($2\,\sigma$) level \cite{HFLAV}.
In the plots, we also put some contours for $|C_X|$ and $\delta_X$ in purple and orange, respectively.
 The constraint from the $B_c^+$ lifetime is shown with the solid black and dashed black lines for $\mathcal{B}(B^+_c \to \tau^+ \nu ) < 0.3$ and $ < 0.1$, respectively.

Note that charged scalar ($H^{\pm}$) scenario gives rise to non-zero $C_{S_{1,2}}$.
A vector boson ($W^{\prime \pm}_L$) that couples to left-handed fermions contributes to $C_{V_1}$.

The vector operators ($O_{V_{1,2}}$) can explain the $R_{D^{(\ast)}}$ anomalies, but $F_{L}^{D^{\ast}}$ has to be the same as the SM prediction ($F_{L,\,\textrm{SM}}^{D^{\ast}} \simeq 0.45$).
For the scalar operators ($O_{S_{1,2}}$), we can see that the constraint from the $B_c^+$ lifetime is significant
and thus the deviations of $R_{D^\ast}$ and $F_{L}^{D^\ast}$ from their SM predictions are severely constrained. 
Finally, $F_L^{D^\ast}$ is suppressed as $R_{D^{(\ast)}}$ are enhanced in the case of the tensor operator $(O_T)$.

\section{Leptoquark scenarios}
\label{sec3}

In this section, we discuss LQ models that can explain the $R_{D^{(\ast)}}$ anomalies,
based on the generic analysis in Sec.~\ref{sec2}.
We address the following three types of LQs with (SU(3$)_C$, SU(2$)_L$, U(1$)_Y)$ SM quantum numbers that are known as good candidates to accommodate the $R_{D^{(\ast)}}$ discrepancies:
\begin{itemize}
\item ${\rm R}_2$ with ($\bf{3},~\bf{2},~\frac{7}{6}$): SU(2$)_L$ doublet scalar LQ\\
The scalar LQ ${\rm R}_2$ can generate significant contributions to $R_{D^{(\ast)}}$ 
(\textit{e.g.,} see Refs. \cite{Tanaka:2012nw,Dorsner:2013tla,Sakaki:2013bfa}).
${\rm R}_2$ does not cause the proton decay since there is no diquark coupling. 
On the other hand, it is known that this scenario is not accessible to the $b\to s \mu\mu$ anomaly at the tree-level.
The loop-level contributions \cite{Becirevic:2017jtw} and the scenario with ${\rm R}_2$--${\rm S}_3$ combination
\cite{Dorsner:2017wwn} have been studied to accommodate both anomalies, where ${\rm S}_3$ with $({\bar {\bf 3}}, {\bf 3}, \frac{1}{3})$  is a SU(2$)_L$ triplet scalar LQ. 
\item ${\rm S}_1$ with ($\bf{\bar{3}},~\bf{1},~\frac{1}{3}$): SU(2$)_L$ singlet scalar LQ \\
The scalar LQ ${\rm S}_1$ is also known as a candidate to explain the $R_{D^{(\ast)}}$ anomalies 
(\textit{e.g.,} see Refs.~\cite{Sakaki:2013bfa,Freytsis:2015qca,Li:2016vvp}). 
In order to ensure the proton stability, we assume that diquark couplings to LQ are forbidden (by a symmetry, see Ref.~\cite{Dorsner:2016wpm}).  
Although this LQ does not provide $b \to s \ell \ell$ transition at the tree-level, the loop-level contributions have been investigated\cite{Bauer:2015knc}.  
Then it is found that the scenario with a pair of ${\rm S}_1$ and ${\rm S}_3$ is viable\cite{Crivellin:2017zlb,Buttazzo:2017ixm,Marzocca:2018wcf}.
\item ${\rm U}_1$ with ($\bf{3},~\bf{1},~\frac{2}{3}$): SU(2$)_L$ singlet vector LQ\\
The ${\rm U}_1$ vector LQ has been receiving attention 
because it can provide a simultaneous explanation of the anomalies in the $b \to s$ and $b \to c$ transitions 
(\textit{e.g.,} see Refs. \cite{Bhattacharya:2016mcc,Buttazzo:2017ixm, Calibbi:2017qbu,Blanke:2018sro,Crivellin:2018yvo}). 
This LQ does not predict the proton decay. 
\end{itemize}

\subsection{Models}
\label{sec:Models}

We adopt the notation of Refs.~\cite{Dorsner:2016wpm,Angelescu:2018tyl}. 
The left-handed doublets are represented as $Q^i = \left ( (V_{\textrm{CKM}}^{\dagger} u_L)^i, \ d_{L}^i \right )^T$ and 
$L^i = \left (  {\nu_{L}^i}, \  \ell_{L}^i \right )^T$, 
where $V_{\textrm{CKM}}$ is the CKM matrix. Here, $u^i $
and $d^i$ denote mass eigenstates. 
Below, we present the relevant couplings in the each model and derive
the effective four-fermi interactions that contribute to the semi-leptonic $B$ decays.

\subsubsection{${\rm R}_2$ LQ model }
We introduce one ${\rm R}_2$ LQ whose SM charges are ($\bf{3},~\bf{2},~\frac{7}{6}$).
 ${\rm R}_2$ is a scalar field, so that it couples to quarks and leptons flavor-dependently via
 Yukawa couplings. The Yukawa interactions involving ${\rm R}_2$ can be written as
\begin{align}
\label{eq:yuk-R2}
\mathcal{L}_{{\rm R}_2} 
= y_R^{ij} \, \overline{Q}_i \ell_{R\,j}\,{\rm R}_2 - y_L^{ij} \, \overline{u}_{R\,i} {{\rm R}_2} i \tau_2 L_j + \mathrm{h.c.}\,,
\end{align}
where $y_L$ and $y_R$ are $3 \times 3$ complex matrices. 
In terms of the electric charge eigenstates, it can be written as
\begin{align}
\label{eq:yuk-R2-bis}
\begin{split}
\mathcal{L}_{{\rm R}_2} 
&= (V_{\textrm{CKM}} y_R)^{ij} \, \overline{u}_{L\,i} \ell_{R\,j}\,{\rm R}_2^{(5/3)} 
+ y_R^{ij} \, \overline{d}_{L\,i} \ell_{R\,j}\,{\rm R}_2^{(2/3)}    \\[0.4em]
&+y_L^{ij} \bar{u}_{R\,i} \nu_{L\,j}\, {\rm R}_2^{(2/3)} 
- y_L^{ij} \overline{u}_{R\,i} \ell_{L\,j}\, {\rm R}_2^{(5/3)} + \mathrm{h.c.}
\end{split}
\end{align}
The superscripts of ${\rm R}_2$ denote the electromagnetic charges of the LQs.
The ${\rm R}^{(2/3)}_2$ exchange gives contributions to $b \to c \tau {\bar \nu_{\tau}}$ at the tree-level, and 
generates the coefficients of the scalar and tensor operators at the scale $\mu=\mu_{\rm LQ}$:

\begin{equation}
C_{S_2} (\mu_{\rm LQ})
= 4 \, C_T(\mu_{\rm LQ}) 
= \dfrac{1}{4 \sqrt{2} G_F V_{cb}} \dfrac{y_{L}^{c \tau}\big{(}y_R^{b \tau}\big{)}^\ast}{m_{R_2}^2 } \, .
\end{equation}
Assuming the Yukawa couplings are aligned to avoid the strong constraints from flavor observables,
sizable $y_{L}^{c \tau}$ and $y_R^{b \tau}$ couplings can achieve the experimental results
of $R_{D^{(\ast)}}$. 
For instance, when one chooses $y_L^{c\tau} (y_R^{b \tau})^{\ast} = 2.5 i $ and $m_{\textrm{LQ}} = 1.5 \ {\rm TeV}$, 
we have $C_{S_2}$ $(4 C_{C_T}) = 0.41 i$ that can explain the present $R_{D^{(\ast)}}$ data within $1\,\sigma$.

\subsubsection{${\rm S}_1$ LQ model} 
Next, we consider a ${\rm S}_1$ LQ whose SM charges are  ($\bf{\bar{3}},~\bf{1},~\frac{1}{3}$). ${\rm S}_1$ is a SU(2)$_L$-singlet scalar, so the Yukawa couplings between ${\rm S}_1$ and the SM fermions can be written as 

\begin{align}
\begin{split}
\mathcal{L}_{{\rm S}_1} 
&= y_L^{ij} \, \overline{Q^C} i\tau_2 L_j\, {\rm S}_1 
+ y_R^{ij} \,\overline{u^C_{R\,i}} e_{R\,j}\, {\rm S}_1 
+\mathrm{h.c.} \\[0.4em]
&= {\rm S}_1 \Big{[}\big{(}V_{\textrm{CKM}}^\ast y_L \big{)}^{ij}\, \overline{u^C_{L\,i}}\ell_{L\,j}-y_L^{ij}\,\overline{d^C_{L\,i}}\nu_{L\,j}+y_R^{ij}\, \overline{u^C_{R\,i}}\ell_{R\,j} \Big{]} + \mathrm{h.c.}\,,
\end{split}
\end{align}

\noindent where $y_L$ and $y_R$ are generic $3 \times 3$ matrices. 
Assuming that $S_1$ is heavy, 
the contribution of the $S_1$ exchange to $b \to c \tau {\bar \nu_{\tau}}$ at the tree-level is given by 
\begin{align}
\label{eq:S1WC}
C_{V_1} (\mu_{\rm LQ})
&= \dfrac{1}{4 \sqrt{2} G_F V_{cb}}\dfrac{y_L^{b \tau}\big{(}V_{\textrm{CKM}} y_L^\ast\big{)}^{c\tau}}{ m_{S_1}^2} \,, \\
C_{S_2} (\mu_{\rm LQ})
&= - 4 \, C_T (\mu_{\rm LQ})
= -\dfrac{1}{4 \sqrt{2} G_F V_{cb}}\dfrac{y_L^{b{\tau}} \big{(}y_R^{c\tau}\big{)}^\ast}{ m_{S_1}^2}\,.
\nonumber
\end{align}
Compared to the ${\rm R}_2$ case, $C_{V_1} $ is also generated.
When one chooses $y_L^{b\tau} (V_{\textrm{CKM}} y_L^\ast)^{c\tau} =0.3,  ~y_L^{b\tau} (y_R^{c\tau})^{\ast} =  -0.3 $ and $m_{\textrm{LQ}} = 1.5 \ {\rm TeV}$, 
$C_{V_1} $ and $C_{S_2} $ $(-4 C_{C_T})$ are $0.05$ at the LQ mass scale,
that can explain the $R_{D^{(\ast)}}$ anomalies at $1\,\sigma$ level.

\subsubsection{${\rm U}_1$ LQ model} 
We also consider a SU(2)$_L$-singlet massive vector LQ, ${\rm U}_1$.
The SM charges of ${\rm U}_1$ are defined as ($\bf{3},~\bf{1},~\frac{2}{3}$).
This LQ is a massive vector field, so that it could be realized by the extension of the SM gauge symmetry. We do not mention the underlying theory, but we simply discuss the phenomenology
introducing flavor-dependent couplings between ${\rm U}_1$ and the SM fermions.
Then, the coupling between ${\rm U}_1$ and the SM fermions can be described by
\begin{equation}
\label{eq:lag-U1}
\mathcal{L}_{{\rm U}_1} 
= x_L^{ij} \, \bar{Q}_i \gamma_\mu {\rm U}_1^\mu L_j 
+ x_R^{ij} \, \bar{d}_{R\,i} \gamma_\mu  {\rm U}_1^\mu \ell_{R\,j}+\mathrm{h.c.},
\end{equation}

\noindent where $x_L^{ij}$ and $x_R^{ij}$ are $3 \times 3$ complex matrices.
Integrating out the heavy ${\rm U}_1$, the couplings contribute to $b\to c\tau \bar{\nu}_{\tau}$ via the following coefficients
\begin{align}
\label{eq:gV-U1}
C_{V_1} (\mu_{\rm LQ})
&= \dfrac{\,\left(V_{\textrm{CKM}} x_L\right)^{c\tau}\left(x_L^{b\tau}\right)^\ast}{ 2 \sqrt{2} G_F V_{cb}\, m_{U_1}^2 },
\\
C_{S_1} (\mu_{\rm LQ})
& =  - \dfrac{\left(V_{\textrm{CKM}} x_L\right)^{c\tau}\left(x_R^{b\tau}\right)^\ast}{\sqrt{2} G_F V_{cb}\, m_{U_1}^2 } .
\label{eq:U1CS1}
\end{align}
Note that the formula of $C_{S_1}$ is omitted in Ref.~\cite{Angelescu:2018tyl}. 
The $C_{V_1}$ contribution interferes with the SM contribution, so that
it easily enhances $R_{D^{(\ast)}}$.  On the other hand, $C_{V_1}$ does not affect
the polarization, as shown in Sec. \ref{sec:operatoranalysis}. 
When 
$(V_{\textrm{CKM}} x_L )^{c\tau}   (x_L^{b \tau})^{\ast} =  0.15$, 
$(V_{\textrm{CKM}} x_L )^{c\tau}   (x_R^{b \tau})^{\ast} =  -0.15$,  and 
$m_{\textrm{LQ}} = 1.5 \ {\rm TeV}$ are taken, $C_{V_1}=0.05$ and $C_{S_1} =  0.1$,  
and the $R_{D^{(\ast)}}$ anomalies can be explained at $1\,\sigma$ level. 
We show our results of the flavor physics in each model, in Sec.~\ref{results}.

\subsection{Renormalization-group running effects}
\label{sec:RGE}

As seen above, some of LQs give rise to contributions from more than two types of the operators. 
In such a case, it is necessary to consider
renormalization-group (RG) evolution effects from the NP scale ($\mu_{\rm NP}$)  to the effective Hamiltonian matching scale ($\mu_b$).
Let us briefly summarize the RG corrections in this subsection.

The semi-leptonic vector and axial vector four-fermion operators do not evolve in QCD~\cite{Buras:1998raa} and there are no operator mixings with the other operators which we consider \cite{Alonso:2013hga}, so that we deal with $O_{V_{1,2}}$ as scale independent operators: $C_{V_{1,2}} (\mu_b) \simeq C_{V_{1,2}} (\mu_{\rm NP})$.

It is pointed out that a large operator mixing between $O_{S_2}$ and $O_T$ arises from the electroweak anomalous dimension \cite{Gonzalez-Alonso:2017iyc} above the electroweak symmetry breaking scale ($\mu_{\rm EW}$).
The RG evolution for the operators  $C_i = \left\{C_{S_1}, C_{S_2}, C_{T} \right\}$
at the one-loop level \cite{Alonso:2013hga, Jenkins:2013wua, Gonzalez-Alonso:2017iyc, Feruglio:2018fxo} 
is given as 
\begin{align}
\frac{d C_{i} (\mu)}{ d \ln \mu} = \frac{1}{16 \pi^2} \left[ g_{s}(\mu)^2 \gamma_{s}^{T} + \gamma_{w}^T(\mu)   + y_t(\mu)^2 \gamma^T_t \right]_{ij}  C_j (\mu),
\label{eq:RGE}
\end{align}
with
\begin{align}
\gamma_{s}^T &= \left\{ \gamma_{S}, \gamma_{S}, \gamma_{T}\right\}_{\rm diag},\\
\gamma_{w}^T(\mu) & = 
\begin{pmatrix}
 - \frac{8}{3} g^{\prime 2} (\mu)  & 0 &0 \\
 0 & - \frac{11}{3} g^{\prime 2}(\mu) & 18 g^2 (\mu) + 30 g^{\prime 2}(\mu)\\
 0 & \frac{3}{8} g^2(\mu) + \frac{5}{8} g^{\prime 2}(\mu) & - 3 g^2(\mu) + \frac{2}{9} g^{\prime 2} (\mu)
\end{pmatrix},\\
\gamma_{t}^T &= \left\{ 0, 1/2, 1/2\right\}_{\rm diag},
\end{align}
where 
$\gamma_S = - 6 C_F = -8$ and $\gamma_T = 2 C_F = 8/3$.
We numerically  solve the RG evolution in Eq.~\eqref{eq:RGE} from $\mu_{\rm NP}$ to $\mu_{\rm EW} = m_Z$ in our analysis.

On the other hand, 
 below the electroweak scale,
the RG evolution is dominated by the QCD contributions.
Reference~\cite{Gonzalez-Alonso:2017iyc}  gives a numerical solution for the RG evolution
at the
three-loop in QCD  and the one-loop in QED as follows,\footnote{
A relation of the operator basis in Ref.~\cite{Gonzalez-Alonso:2017iyc} with our basis is 
\begin{align}
\begin{pmatrix}
\epsilon_S \\
\epsilon_P \\
\epsilon_T
\end{pmatrix}
= 
\begin{pmatrix}
1 & 1 & 0\\
-1 & 1 & 0 \\
0 & 0 & 1
\end{pmatrix}
\begin{pmatrix}
C_{S_1} \\
C_{S_2} \\
C_T
\end{pmatrix}.
\end{align}
}
\begin{align}
\begin{pmatrix}
C_{S_1} (\mu_b) \\
C_{S_2} (\mu_b) \\
C_T (\mu_b) 
\end{pmatrix}
\simeq 
\begin{pmatrix}
1.46 & 0 &  0\\
0 & 1.46 & -0.0177 \\
 0 & -0.0003 & 0.878
\end{pmatrix}
\begin{pmatrix}
C_{S_1} (m_Z) \\
C_{S_2} (m_Z) \\
C_T (m_Z)
\end{pmatrix} .
\end{align}

When one considers the RG evolution at the three-loop level in QCD and the one-loop level in electroweak and QED mentioned above, 
one obtains  \cite{Gonzalez-Alonso:2017iyc} 
$C_{S_2}(\mu_b) \simeq + 8.1 C_{T}(\mu_b)$ [$C_{S_2}(\mu_b) \simeq -8.5 C_{T}(\mu_b)$] 
when $C_{S_2}(\mu_\text{NP}) = + 4 C_{T}(\mu_\text{NP})$ [$C_{S_2}(\mu_\text{NP}) = - 4 C_{T}(\mu_\text{NP})$] at $\mu_\text{NP} = \mathcal O(1)\text{TeV}$. 
Therefore, the ratio of $C_{S_2}(\mu_{b})$ to $C_{T}(\mu_b)$ in the case with 
$C_{S_2}(\mu_{\rm NP}) = - 4 C_{T}(\mu_{\rm NP})$ is more amplified than in the case with $C_{S_2}(\mu_{\rm NP}) = + 4 C_{T}(\mu_{\rm NP})$ at $\mu_\text{NP} = \mathcal O(1)\text{TeV}$.\footnote{
On the other hand, if one considers the RG evolution from $\mu_{\rm NP}$ to $\mu_{b}$ at only the one-loop level in QCD, there is no operator mixing and the exact solution is given as \cite{Dorsner:2013tla,Sakaki:2013bfa},
\begin{align}
C_{S_{i}} (\mu_b ) &= \left[ \frac{\alpha_s (m_t)}{\alpha_s (\mu_b)}\right]^{\frac{\gamma_S}{2 \beta_0^{(5)}}}\left[ \frac{\alpha_s (\mu_{\rm NP})}{\alpha_s (m_t)}\right]^{\frac{\gamma_S}{2 \beta_0^{(6)}}} C_{S_{i}} (\mu_{\rm NP}) \textrm{~~~for~}i=1,2, \\
C_{T} (\mu_b) &= \left[ \frac{\alpha_s (m_t)}{\alpha_s (\mu_b)}\right]^{\frac{\gamma_T}{2 \beta_0^{(5)}}}\left[ \frac{\alpha_s (\mu_{\rm NP})}{\alpha_s (m_t)}\right]^{\frac{\gamma_T}{2 \beta_0^{(6)}}} C_{T} (\mu_{\rm NP}),
\end{align}
where $\beta_0^{(f)} =  11 - 2f/3$.
Then, one obtains 
$C_{S_2}(\mu_b) \simeq \pm 7.7 C_{T}(\mu_b)$ when $C_{S_2}(\mu_{\rm NP}) = \pm 4 C_{T}(\mu_{\rm NP})$ holds. 
}


\subsection{Results}
\label{results}
Here, we discuss whether $F_L^{D^{\ast}}$ could be enhanced in the LQ scenarios that can accommodate the current $R_{D^{(\ast)}}$ anomalies. 
Note that the present case is different from the scenarios with the single NP operator (see Sec.~\ref{sec:operatoranalysis}) 
in the sense that various NP operators are induced from the LQ interactions and thus contributions to the observables are non-trivial.

\begin{figure}[tp]
  \begin{center}
  \subfigure[${\rm R}_2$ LQ]
    {\includegraphics[width=0.48\textwidth]{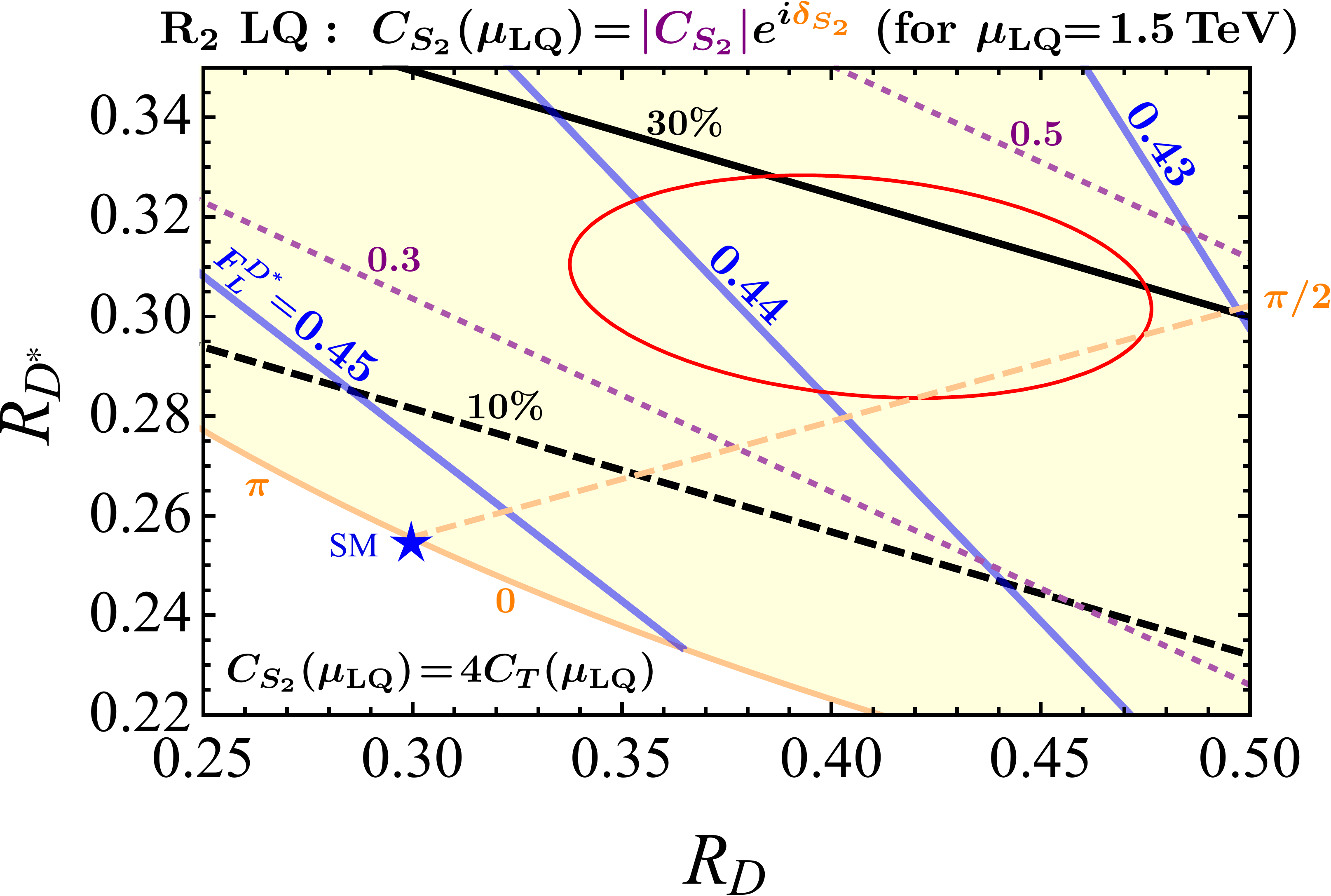} \label{fig:R2}}\\
       \vspace{0.2cm}
  \subfigure[${\rm S}_1$ LQ with $C_{V_1} (\mu_{\textrm{LQ}})= 0$]
    {\includegraphics[width=0.48\textwidth]{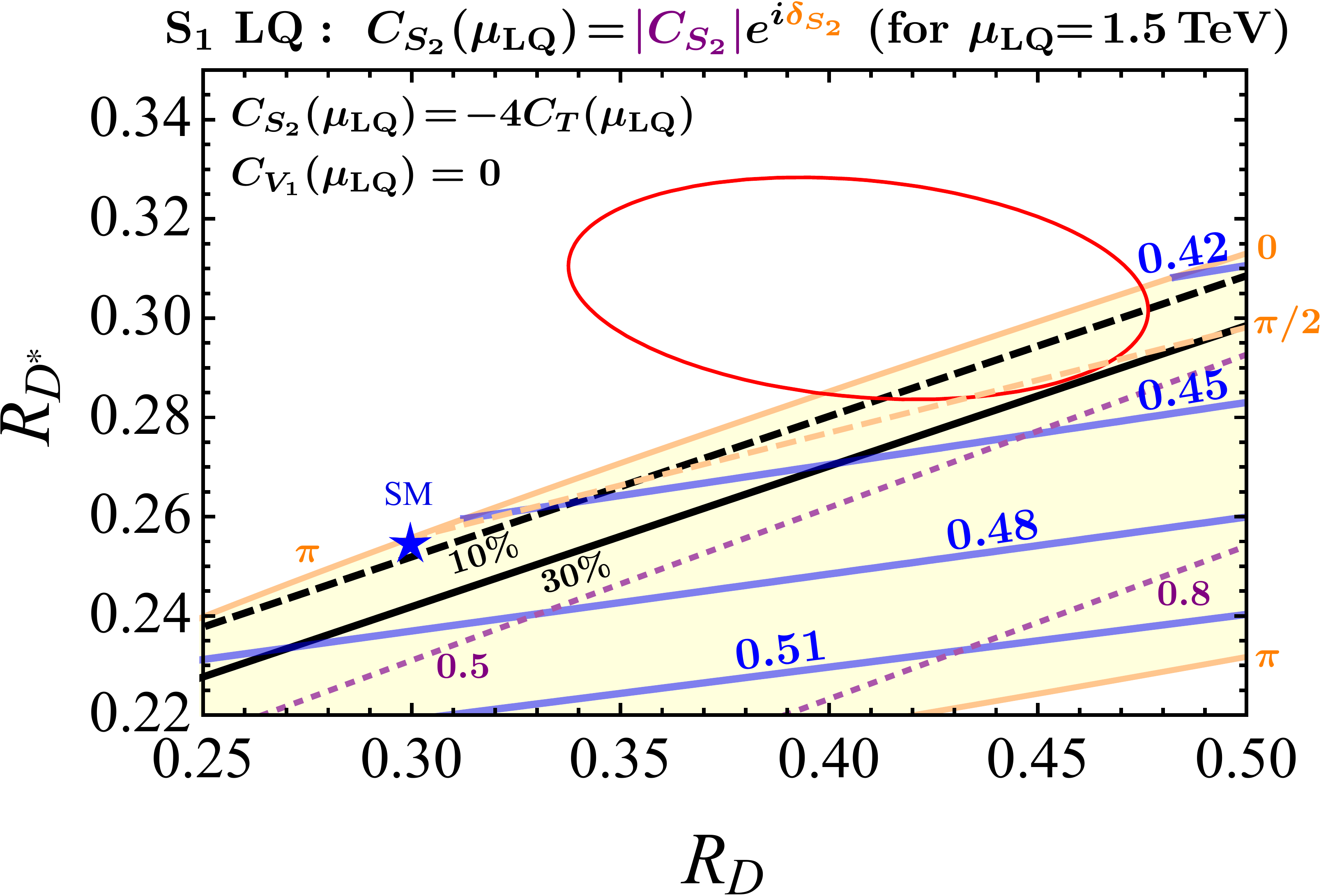}  \label{fig:S1}}
      \subfigure[${\rm S}_1$ LQ with $C_{V_1} (\mu_{\textrm{LQ}})= 0.1$]
    { \includegraphics[width=0.48\textwidth]{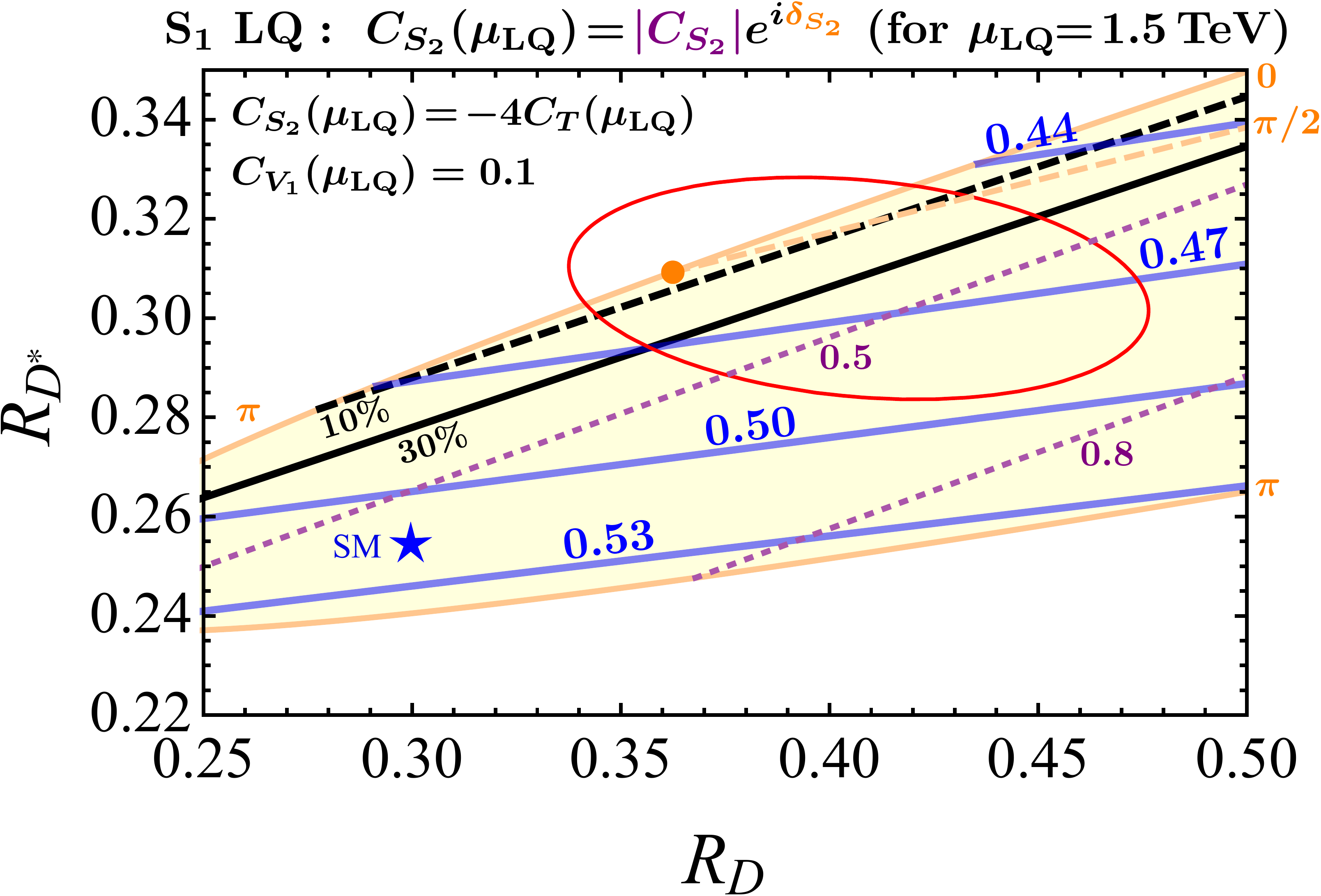} \label{fig:S12}}
   \\
   \vspace{0.2cm}
  \subfigure[${\rm U}_1$ LQ with $C_{V_1} (\mu_{\textrm{LQ}})= 0.1$]
    {\includegraphics[width=0.48\textwidth]{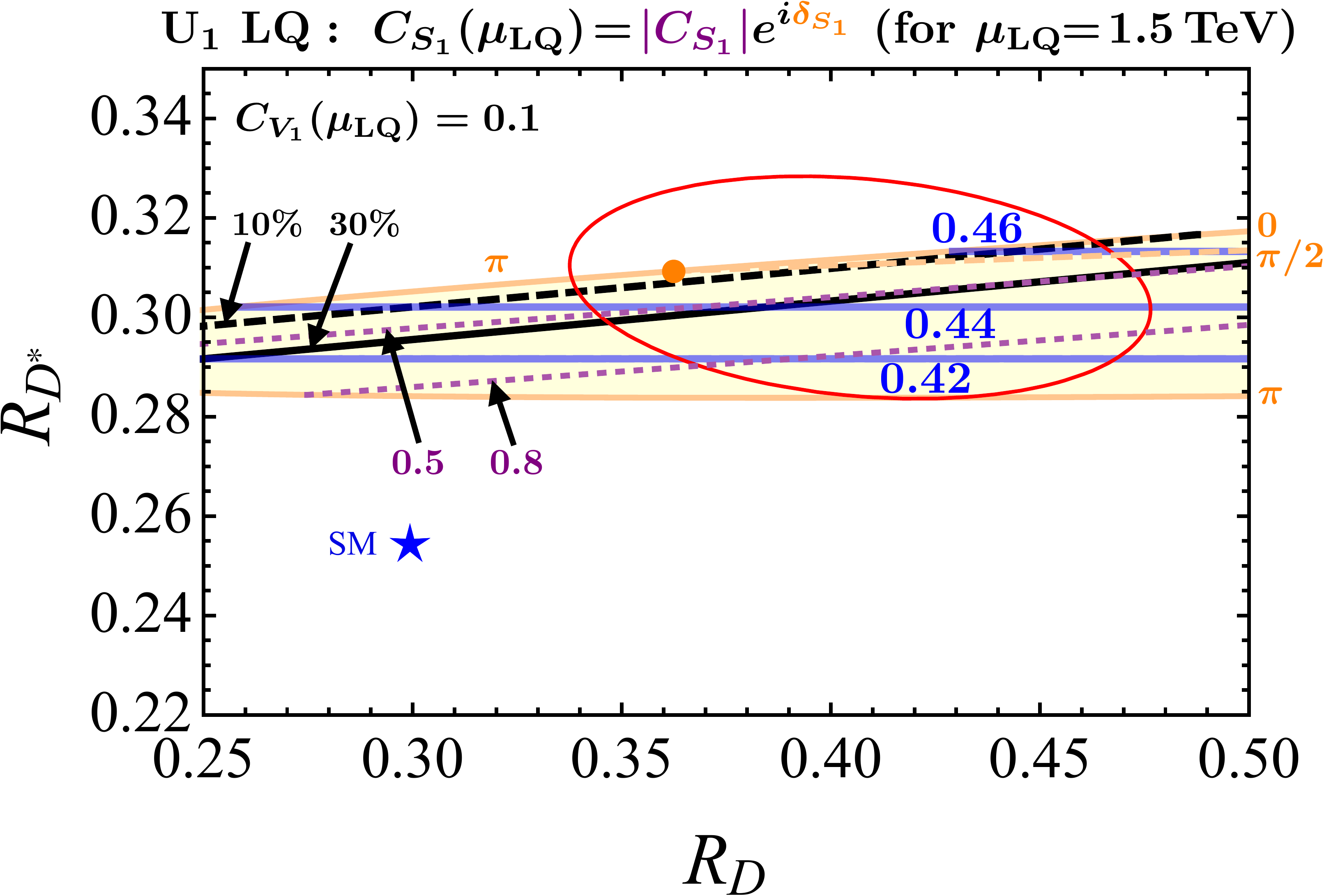}  \label{fig:U1}}
    \caption{
    The $F_L^{D^{\ast}}$ contour is shown on the $R_D$--$R_{D^{\ast}}$ plane with the blue line in 
    (a) ${\rm R}_2$, (b) ${\rm S}_1$ with $C_{V_1} =0$, (c) ${\rm S}_1$  with $C_{V_1} = 0.1$ and (d) ${\rm U}_1$ with $C_{V_1} = 0.1$ LQ scenarios at $m_{\rm LQ} = 1.5 \ {\rm TeV}$.
    The plot legend is the same as that in Fig.~\ref{fig:oneoperator}.
    The orange point stands for the case of $C_{S_2}=0$  and $C_{V_1}=0.1 $ for the $\textrm{S}_1$ LQ or $C_{S_1}=0$ and $C_{V_1}=0.1 $ for the $\textrm{U}_1$ LQ.
    }
    \label{fig:LQ}
  \end{center}
\end{figure}

In Fig.~\ref{fig:LQ}, the $F_L^{D^{\ast}}$ contour is shown on the $R_D$--$R_{D^{\ast}}$ plane with the blue line in the three LQ models: ${\rm R}_2$, ${\rm S}_1$ and ${\rm U}_1$. 
We take $m_{\rm LQ}=1.5 \,{\rm TeV}$ for a reference value of the LQ mass in our analysis, 
where the value is chosen so that the recent collider bounds are satisfied, {\it e.g.}, see Ref.~\cite{Angelescu:2018tyl} for a review. 
Note that the LQ mass is relevant to the RG evolution effects and thus indicated in the plots. 
Then, the correlations among $R_D$, $R_{D^{\ast}}$, and $F_L^{D^{\ast}}$ are seen when varying the Wilson coefficients $C_{X} (\mu_{\rm LQ})$ in the complex plane.
Here, we assume that the couplings of $y_{L,R}$ and $x_{L,R}$, relevant to $\Bb \to D^{(\ast)} \tau \overline{\nu}$, are sizable while the others are negligible in our analysis.
In these plots, we focus on the parameter region that is favored by the $R_{D^{(\ast)}}$ experimental results.\footnote{
The $R_{J/\psi}$ anomaly would overshoot the $R_{D^{\ast}}$ preferred region \cite{Watanabe:2017mip}.
} 

Along with the $F_L^{D^{\ast}}$ contour, we also show the lines for the absolute value and phase of $C_{S_{1,2}} (\mu_{\rm LQ}) \equiv |C_{S_{1,2}}| e^{i \delta_{S_{1,2}}}$ in purple and orange, respectively, in the figures. 
Note that the shaded region in yellow can be achieved in the single LQ scenario.
The constraint from the $B_c^+$ lifetime is put with the solid (dashed) black lines corresponding to $\mathcal{B}(B^+_c \to \tau^+ \nu ) = 0.3\ (0.1)$. 
The SM point and the present data for $R_{D^{(\ast)}}$ are denoted by the blue star and the red ellipse, respectively.
 The orange points stand for the cases of $C_{S_2}=0$ for the $\textrm{S}_1$ LQ and $C_{S_1}=0$ for the $\textrm{U}_1$ LQ.

In the ${\rm R}_2$ LQ model, the scalar Wilson coefficient $C_{S_2}$ and tensor one $C_T$ are introduced. 
In the Fig.~\ref{fig:R2}, it is found that $F_L^{D^{\ast}}$ is not so changed from the SM point ($F_{L,\,\text{SM}}^{D^{\ast}} \simeq 0.45$) in this scenario.
Our prediction of a range of $F_L^{D^{\ast}}$ within the present data of $R_{D^{(\ast)}}$ at $1\,\sigma$, and within the $B_c^+$ lifetime bound [$\mathcal{B}(B_c^+ \to \tau^+ \nu) <0.3$], is  $[0.43, 0.44]$.
The value of $R_{D^{(\ast)}}$ is constrained by the $B_c^+$ lifetime. 
If we take $\mathcal{B}(B_c^+ \to \tau^+ \nu) <0.3$, $R_{D^{(\ast)}}$ is loosely constrained: 
the present data is still accommodated (with $|C_{S_2}(\mu_{{\rm LQ}})| = 4|C_T(\mu_{{\rm LQ}})| \sim 0.4$ in the vicinity of $\delta_{S_2} = \pi/2$, for instance). 
The result is consistent with Refs.~\cite{Dorsner:2013tla,Sakaki:2013bfa,Hiller:2016kry,Becirevic:2018afm}.

In the ${\rm S}_1$ LQ model, the $C_{V_1}$, $C_{S_2}$, and $C_{T}$ operators are introduced with the relation as $C_{S_2}(\mu_{{\rm LQ}}) = -4C_T(\mu_{{\rm LQ}})$.
The phase of $C_{V_1}$ can be  absorbed by the redefinition of $C_{S_2,T}$ as shown in Appendix \ref{phasedorp}, and thus only three parameters remain: $\delta_{S_2}$, $|C_{S_2}|$, and $|C_{V_1}|$.
(Note that $C_{V_1}$ and $C_{S_2 (T)}$ can be independent by using $y_L^{i\tau}$ and $y_R^{c \tau}$.) 
Then, the $F_L^{D^{\ast}}$ contour is shown for the cases of  $|C_{V_1}|=0$ and $|C_{V_1}|=0.1$ in Fig.~\ref{fig:S1} and Fig.~\ref{fig:S12}, respectively.
It is found that a large $F_L^{D^{\ast}}$ is disfavored by the $B_c^+$ lifetime. 
The case for $|C_{V_1}|=0$ cannot explain the central value of the present $R_{D^{\ast}}$ data while that for $|C_{V_1}|=0.1$ can do~\cite{Cai:2017wry}. 
We can see that the constraint $\mathcal{B}(B_c^+ \to \tau^+ \nu) <0.3$ is satisfied for the latter case. 
Finally, varying the value of $|C_{V_1}|$ we find that the ${\rm S}_1$ scenario predicts a range of $F_L^{D^{\ast}}$ as $[0.42, 0.48]$.

In the ${\rm U}_1$ LQ model, the relevant Wilson coefficients are $C_{S_1}$ and $C_{V_1}$. 
In the same way as the ${\rm S}_1$ LQ, $|C_{V_1}|$, $|C_{S_1}|$, and $\delta_{S_1}$ are free parameters. 
The result for $|C_{V_1}|=0$ is the same as the one in one operator analysis $C_{S_1}$ in Fig.~\ref{fig:oneoperator}.
The case for $|C_{V_1}|=0.1$ is shown in the Fig.~\ref{fig:U1}.
We find that the ${\rm U}_1$ LQ predicts a range of $F_L^{D^{\ast}}$ as $[0.43, 0.47]$ and is consistent with the present data of $R_{D^{(\ast)}}$ and the bound of $\mathcal{B}(B_c^+ \to \tau^+ \nu) <0.3$.

\begin{figure}[tp]
  \begin{center}
  \subfigure[${\rm R}_2$ LQ]
    {\includegraphics[width=0.48\textwidth]{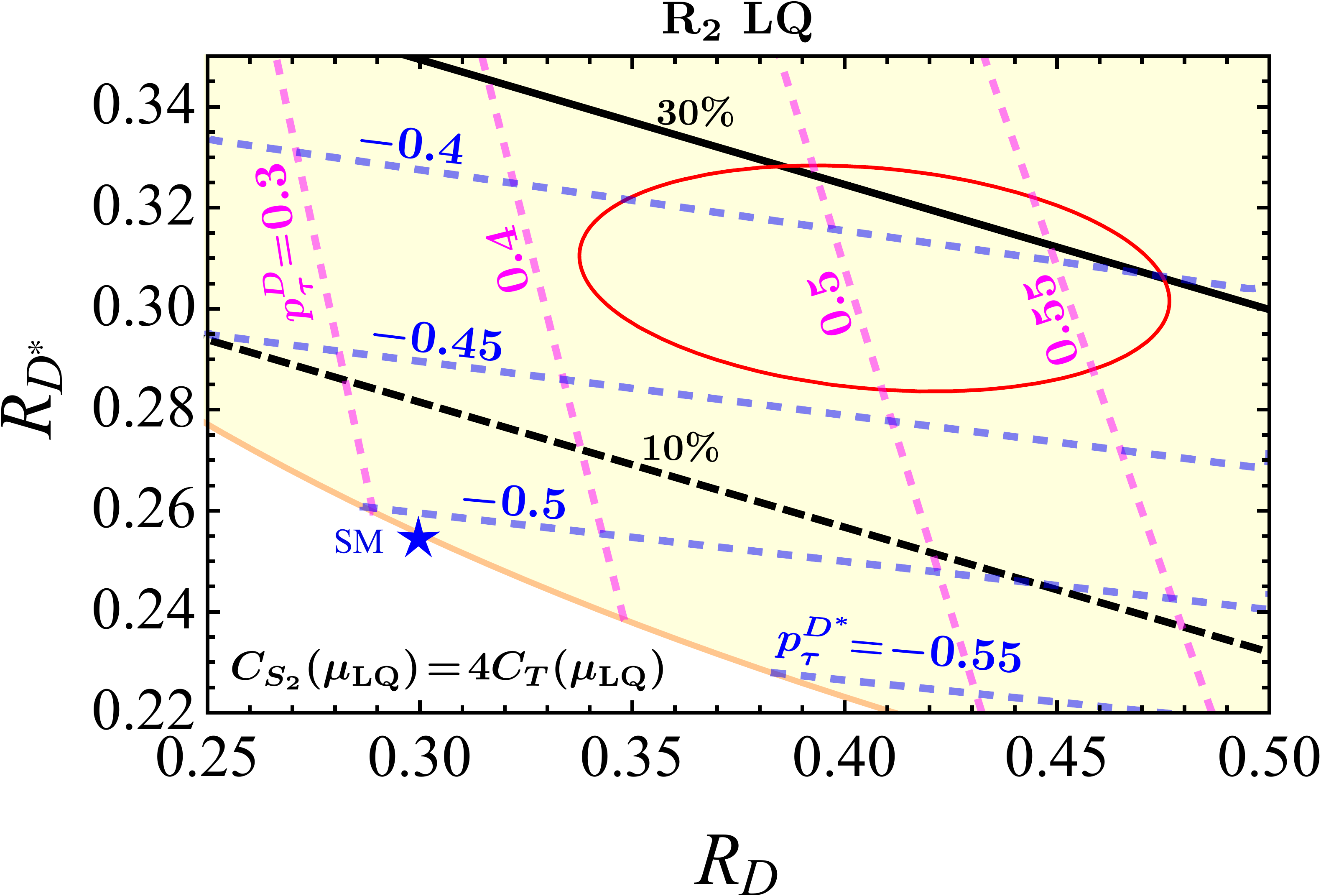} \label{fig:R2tau}}\\
       \vspace{0.2cm}
  \subfigure[${\rm S}_1$ LQ with $C_{V_1} (\mu_{\textrm{LQ}})= 0$]
    {\includegraphics[width=0.48\textwidth]{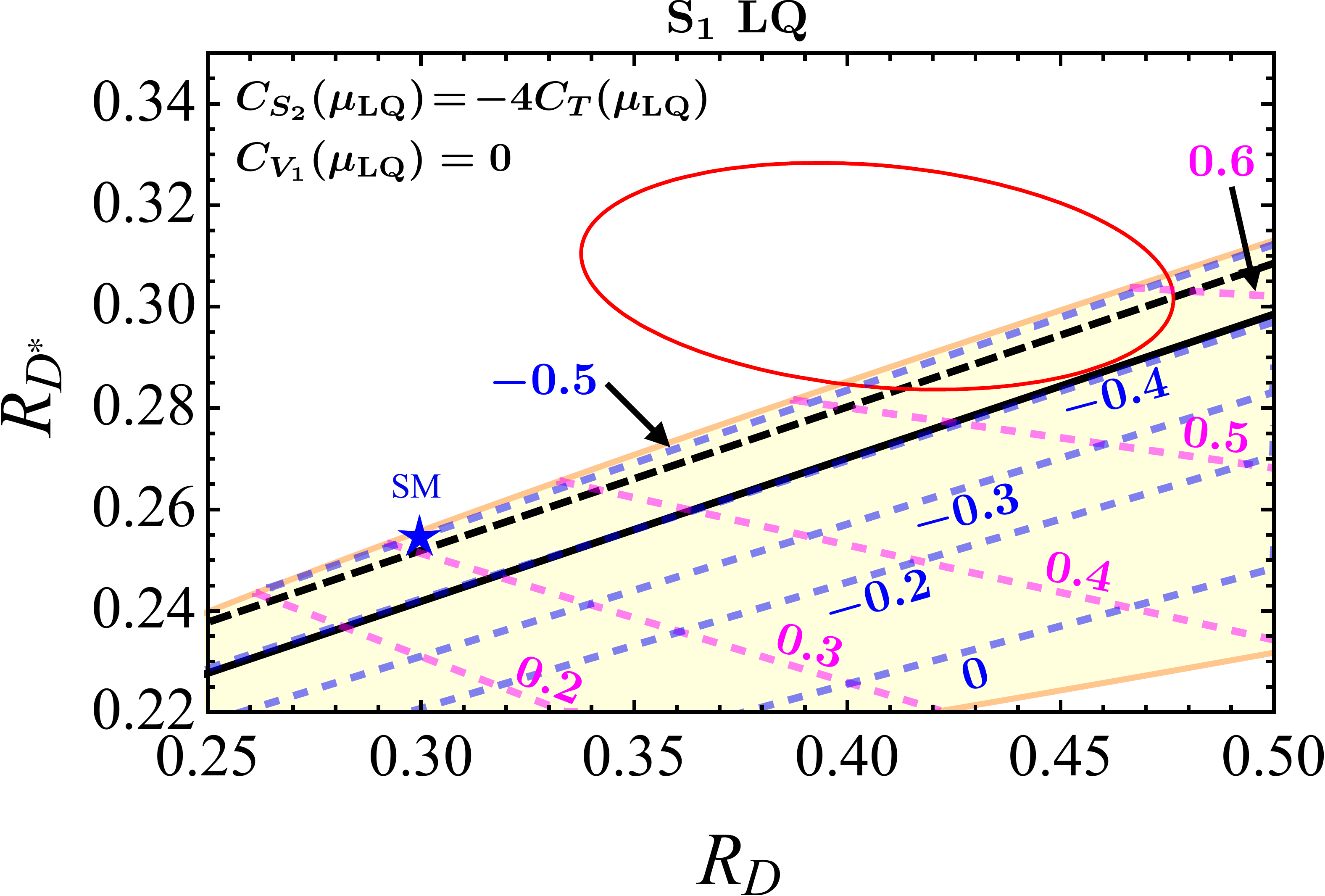}  \label{fig:S1tau}}
      \subfigure[${\rm S}_1$ LQ with $C_{V_1} (\mu_{\textrm{LQ}})= 0.1$]
    { \includegraphics[width=0.48\textwidth]{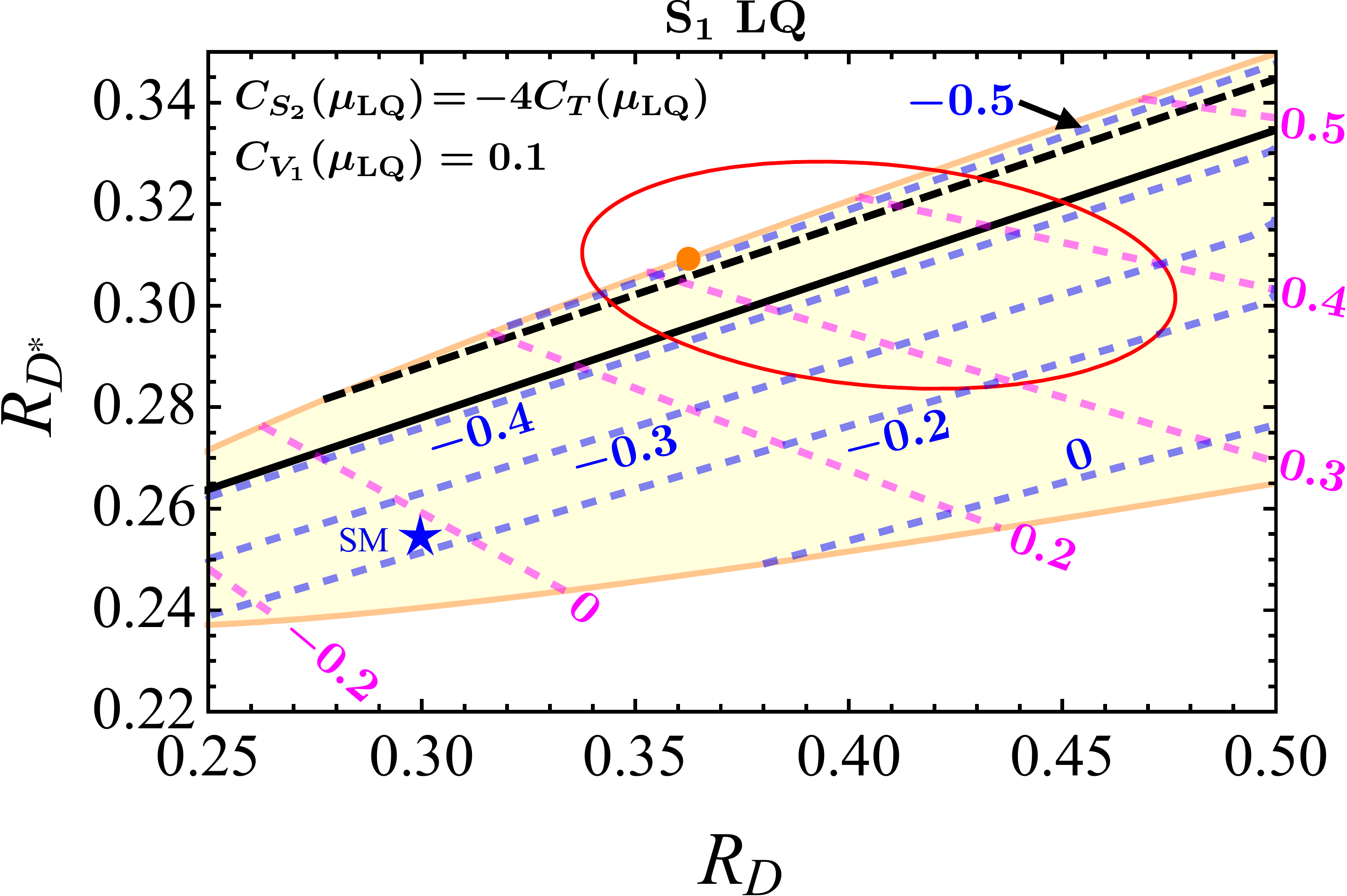} \label{fig:S1tau2}}
   \\
   \vspace{0.2cm}
  \subfigure[${\rm U}_1$ LQ with $C_{V_1} (\mu_{\textrm{LQ}})= 0.1$]
    {\includegraphics[width=0.48\textwidth]{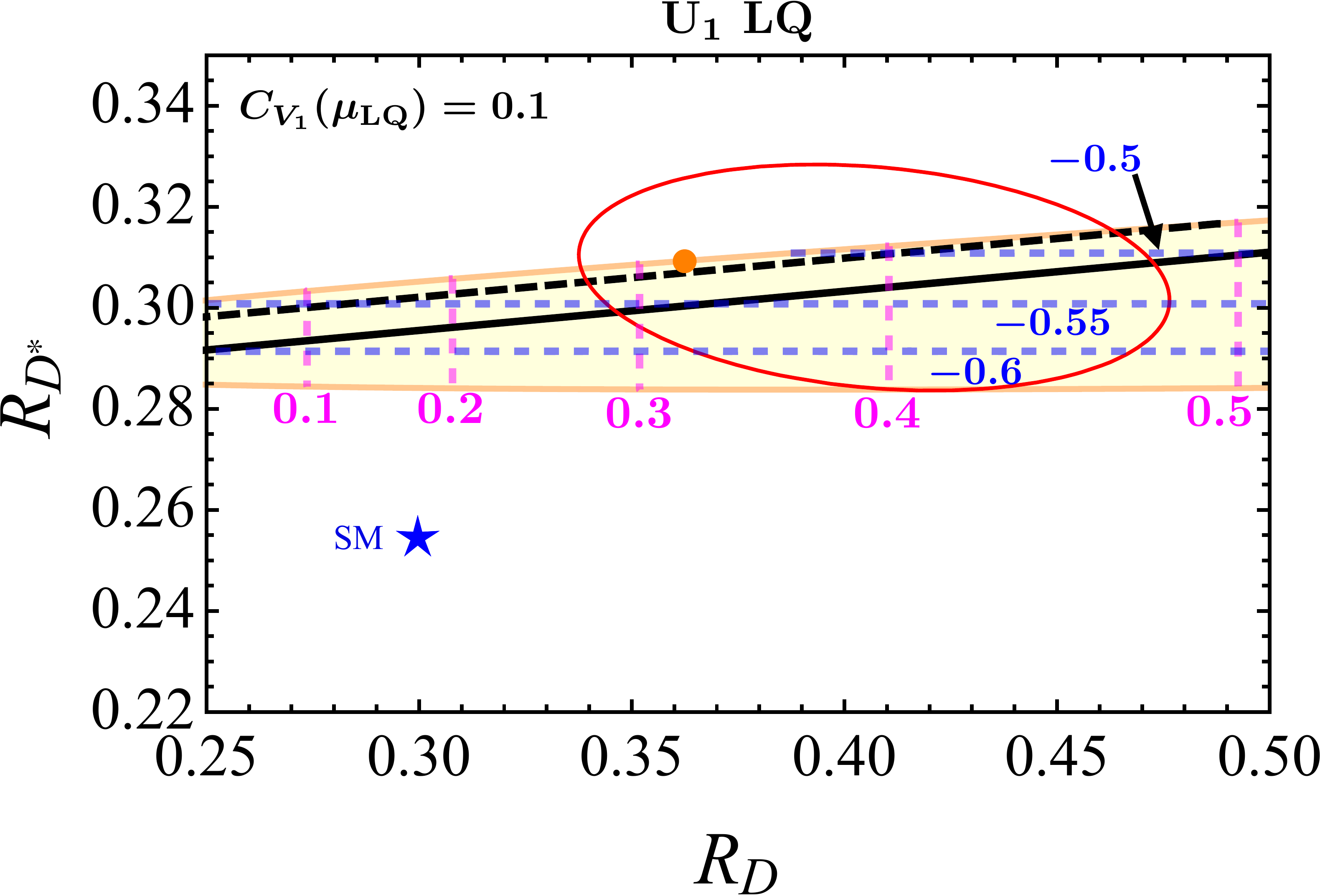}  \label{fig:U1tau}}
    \caption{
   The contours of the $\tau$ polarizations $P_{\tau}^{D}$ and $P_{\tau}^{D^{\ast}}$ are shown on the $R_D$--$R_{D^{\ast}}$ plane in magenta and blue colors, respectively, when we consider
    the LQ scenarios of 
    (a) ${\rm R}_2$, (b) ${\rm S}_1$  with $C_{V_1} = 0$, 
    (c) ${\rm S}_1$  with $C_{V_1}= 0.1$  and (d) ${\rm U}_1$ with $C_{V_1} = 0.1$.
    In all cases, $m_{\rm LQ} = 1.5 \ {\rm TeV}$ is taken.
    The SM point is represented by the blue star.
        }
    \label{fig:LQtaupolari}
  \end{center}
\end{figure}

In the end, we found that 
these three LQ models cannot give $F_L^{D^{\ast}}$ deviating from the SM prediction ($F_{L,\,\text{SM}}^{D^{\ast}} \simeq 0.45$) as long as we take the present $R_{D^{(\ast)}}$ data seriously, especially 
a large deviation of $F_L^{D^{\ast}}$ is restricted by the severe constraint from the $B_c^+$ lifetime in the ${\rm S}_1$ and ${\rm U}_1$ models. 
Figure~\ref{fig:LQ} shows that the LQ models can not explain the experimental result for $F_L^{D^{\ast}}$ in Eq.~\eqref{eq:FLDst_exp} at $1\,\sigma$ level. 
On the other hand, the large enhancement of $R_{D^{(\ast)}}$, 
compared with the SM predictions, is still possible 
although the severe constraint from the $B_c^+$ lifetime excludes some regions of the parameter space. 
Therefore, 
the large/small deviation of $R_{D^{(\ast)}}$/$F_L^{D^{\ast}}$ is one of the possibilities in the LQ models, 
which will be verified at the Belle II experiment. 
If this is the case, however, it is difficult to distinguish the LQ scenarios.

In turn, we study correlation between $R_{D^{(\ast)}}$ and the $\tau$ polarizations $P_{\tau}^{D^{(*)}}$.
In Fig.~\ref{fig:LQtaupolari}, the contours of $P_{\tau}^D$ and $P_{\tau}^{D^{*}}$ are shown with dashed lines in magenta and blue, respectively.
The other legends in the plots are the same as Fig.~\ref{fig:LQ}.
We can see that each LQ model predicts unique ranges for $P_{\tau}^D$ and $P_{\tau}^{D^{\ast}}$, which can be used to distinguish these LQ models:
$( P_{\tau}^D,\, P_{\tau}^{D^{\ast}} )$ with  
([$0.42$, $0.57$],~[$-0.44$, $-0.39$]) for ${\rm R}_2$ LQ, 
([$0.11$, $0.63$],~[$-0.51$, $-0.41$]) for ${\rm S}_1$ LQ and 
([$0.23$, $0.52$],~[$-0.57$,$-0.47$]) for ${\rm U}_1$ LQ 
are predicted where the current data of $R_{D^{(*)}}$ at $1\,\sigma$
and the bound of $\mathcal{B}(B_c^+ \to \tau^+ \nu) <0.3$ are satisfied.
Here, $C_{V_1} (\mu_{\textrm{LQ}}) $ is also varied in ${\rm S}_1$ and ${\rm U}_1$ LQ models.
Note that the predicted ranges of $P_{\tau}^{D^{\ast}}$ are consistent with the latest result by the Belle experiment \cite{Hirose2016wfn, Hirose:2017dxl}
\beq
P_{\tau}^{D^{\ast}} = - 0.38 \pm 0.51(\textrm{stat.}){}^{+0.21}_{-0.16}(\textrm{syst.}).
\eeq
Since Belle~II with $50\,{\rm ab} ^{-1}$ data can measure $P_{\tau}^D$ with $3 \%$ accuracy \cite{Alonso:2017ktd},\footnote{
Only statistical uncertainty has been considered  \cite{Alonso:2017ktd}. 
} and $P_{\tau}^{D^{\ast}}$ with $\pm 0.07$ \cite{Kou:2018nap}, 
we point out that the future measurement of $P_{\tau}^D$ has sufficient sensitivity to distinguish between the LQ models. 
Note that $W^{\prime}$ models predict $P_{\tau}^D = P_{\tau,\,\textrm{SM}}^D$ for any values of $C_{V_1}$ and $C_{V_2}$. 
Thus, $P_{\tau}^D$ is a good observable for discrimination between  $W^{\prime}$ and LQ models.

In Table~\ref{tab:summary}, we summarize our results of the predictions on the polarization observables  for the LQ models. 
This can be partly compared with Ref.~\cite{Hu:2018veh} based on the SM effective field theory.
Note that the uncertainties for the SM predictions are taken from Refs.~\cite{Bernlochner:2017jka,Alok:2016qyh,Hirose2016wfn}. 
We also stress that our study provides the theoretically possible ranges of the polarization observables which satisfy the current $R_{D^{(\ast)}}$ data at $1\,\sigma$ level,
by scanning the full set of the parameters in the LQ models. 
On the other hand, model-independent and -dependent parameter fits from the data including $F_L^{D^{\ast}}$ are performed in Refs.~\cite{Aebischer:2018iyb, Blanke:2018yud}.

\begin{table}[t!]
\begin{center}
  \caption{
  Predicted ranges of the polarizations for ${\rm R}_2$, ${\rm S}_1$ and ${\rm U}_1$ LQ models ($\mu_{\textrm{LQ}} = 1.5$\,TeV), 
  which satisfy the current $1\,\sigma$ data of $R_{D^{(*)}}$ and the bound of $\mathcal{B}(B_c^+ \to \tau^+ \nu) <0.3$. 
  The SM predictions, the current data, and the expected sensitivity at Belle~II with $50\,{\rm ab} ^{-1}$ data \cite{Alonso:2017ktd,Kou:2018nap,Adamczyk:2019wyt} are also shown.
  The sensitivities  for $F_{L}^{D^{\ast}}$ and $P_{\tau}^{D^{\ast}}$ are absolute uncertainty while the others are relative.
}
\label{tab:summary}
\small{
\begin{tabular}{c | ccccc}
\hline
\hline 
 & $ F_{L}^{D^{\ast}}$ & $P_{\tau}^D$ & $P_{\tau}^{D^{\ast}}$ &$R_{D}$ & $R_{D^{\ast}}$  \\
\hline
${\rm R}_2$ LQ & [$0.43$, $0.44$] & [$0.42$, $0.57$] & [$-0.44$, $-0.39$] & $1\,\sigma$ data & $1\,\sigma$ data \\
${\rm S}_1$ LQ & [$0.42$, $0.48$] & [$0.11$, $0.63$] & [$-0.51$, $-0.41$] & $1\,\sigma$ data &$1\,\sigma$ data\\
${\rm U}_1$ LQ & [$0.43$, $0.47$] & [$0.23$, $0.52$] & [$-0.57$,$-0.47$] & $1\,\sigma$ data & $1\,\sigma$ data\\
 \hline
SM & $ 0.46(4)$ & $0.325(9)$ & $-0.497(13)$ & $0.299(3)$ & $0.258(5)$\\
\hline
data & $ 0.60(9)$ & - & $ -0.38(55)$ & $0.407(46)$ & $0.306(15)$\\
Belle II & $0.04$  & $3\%$ & $0.07$ & $3\%$ & $2\%$\\
\hline \hline
\end{tabular}
}
\end{center}
\end{table}

\clearpage
Before closing this section, we comment on the LQ mass dependence.
Since there is no operator mixing, the figures for ${\rm U}_1$ LQ are independent of the LQ mass scale. 
Predicted ranges of $F_L^{D^{\ast}}$ and $P_{\tau}^{D^{(\ast)}}$ slightly depend on the LQ mass scale
through the electroweak RG evolution in ${\rm R}_2$ and  ${\rm S}_1$  LQ cases (see Sec.~\ref{sec:RGE}).  
We found that the variations of  $F_L^{D^{\ast}}$ and $P_{\tau}^{D^{(\ast)}}$  are at the most $0.01$ when $1\,$TeV $< \mu_{\textrm{LQ}} < 3\,$TeV is taken.

\section{Conclusion}
\label{conclusion}

The observed excesses of $R_{D^{(\ast)}}$ in $\Bb \to D^{(\ast)} \tau \overline{\nu}$ have been one of the major anomalies in particle physics since the combined deviation is $3.8\,\sigma$ at present. 
Thus, it is important to summarize the NP explanations and investigate how to hunt the NP footprint.  
There are several ways to test the NP predictions according to the direct and indirect searches for NP signals.
In fact, it is found that the $B_c^+ $ lifetime severely constrains the NP scenarios, even though the leptonic decay, $B_c^+ \to \tau^+ \nu$, is still not directly observed.
Moreover, it is recently pointed out that the direct search for the heavy resonance almost excludes  
the charged scalar scenario \cite{Iguro:2018fni}.
The measurements of the physical observables in $\Bb \to D^{(\ast)} \tau \overline{\nu}$
could also conclude the NP possibilities, 
as discussed in Refs.~\cite{Tanaka:2010se,Tanaka:2012nw,Sakaki:2013bfa,Alonso:2016gym,Alok:2016qyh,Ivanov:2017mrj,Alonso:2017ktd}. 
Recently, the Belle collaboration has reported the new result on the longitudinal $D^{\ast}$ polarization $F_{L}^{D^{\ast}}$ in $\Bb \to D^{\ast} \tau \overline{\nu}$, which could give us a new hint about the NP sector.

In this paper, we have investigated the correlations between the ratio $R_{D^{(\ast)}}$ and the $D^*$ polarization $F_{L}^{D^{\ast}}$ for the LQ models in terms of the general effective Hamiltonian. 
It is already known that the three types of LQs can easily explain the present $R_{D^{(\ast)}}$ anomalies; scalar LQs (${\rm R}_2$, ${\rm S}_1$) and vector LQ (${\rm U}_1$).
Since the recent Belle result ($F_{L}^{D^{\ast}} = 0.60 \pm 0.09$) is slightly above the SM prediction ($F_{L,\,\text{SM}}^{D^{\ast}} \simeq 0.45$), 
the NP effect that enhances $F_{L}^{D^{\ast}}$ tends to be favored, which is not achievable with the single NP operators. 
Thus, we have tried to see if this could be possible in the LQ models that induce various types of NP operators.
We, however, conclude that the possible deviations of $F_{L}^{D^{\ast}}$ from the SM prediction are small in the three LQ models. 
We find the predicted ranges of $F_{L}^{D^{\ast}}$ in the LQ models:  
$[0.43, 0.44]$, $[0.42, 0.48]$, and $[0.43, 0.47]$  for ${\rm R}_2$, ${\rm S}_1$ and ${\rm U}_1$, respectively, 
in which the present $R_{D^{(\ast)}}$ anomaly can be explained within $1\,\sigma$.
To be precise, 
it is found that $\mathcal{B}(B_c^{+} \to \tau^{+} \nu)$ severely restricts deviation of $F_{L}^{D^{\ast}}$ from the SM prediction in the ${\rm S}_1$ and ${\rm U}_1$ LQ models. 
In the ${\rm R}_2$ LQ case, $F_{L}^{D^{\ast}}$ is not much influenced. 
Therefore, it is unlikely to accommodate the present data of $R_{D^{(\ast)}}$ and $F_{L}^{D^{\ast}}$ simultaneously at $1\,\sigma$.

We also investigated the correlations between the $R_{D^{(\ast)}}$  explanation and the $\tau$ polarization asymmetries $P_{\tau}^{D^{(*)}}$ in the LQ models.
It is found that the $\tau$ polarization observables can much deviate from the SM predictions.

In Table~\ref{tab:summary}, predicted ranges of the polarizations for the LQ models are summarized. 
Then, we would point out that the upcoming Belle~II experiment can survey the correlations between $R_{D^{(\ast)}}$ and the polarization observables 
with high accuracy enough to discriminate among the LQ models. 
Note that LHCb run~II will also improve the $R_{D^{\ast}}$ observation~\cite{Bediaga:2018lhg}.
According to our results, one can comment on
 a potential for future measurements of the polarization observables.
  As aforementioned in Sec.~\ref{sec:intro}, 
  the present data of $F_L^{D^{\ast}}$ is more than $1\,\sigma$ away from the SM prediction, although it still includes not small uncertainties. 
  At present, 
  the statistical error is dominant 
  and 
  this can be improved in the future measurement at the Belle~II experiment \cite{Adamczyk:2019wyt}.  
  Provided that the present systematic error still remains, 
it is found that  ${\rm R}_2$,  ${\rm S}_1$, and  ${\rm U}_1$ LQ models are excluded  
at the  $3\,\sigma$ level,
   if the present central value ($F_L^{D^{\ast}} \simeq 0.6$) is not changed. 
   In the case, on the other hand, that data becomes consistent with the SM prediction ($F_L^{D^{\ast}} \simeq 0.46$), 
    correlations between the other observables, $R_{D^{(\ast)}}$ and $P_{\tau}^{D^{(\ast)}}$, are significant to probe NP effects.

\section*{Acknowledgements}
We would like to thank Ivan Ni\v{s}and\v{z}i\'c and  Olcyr Sumensari for the numerical comparison with their codes.
We are also grateful to 
Stefan de Boer, Gino Isidori, Satoshi Mishima, Minoru Tanaka, Kazuhiro Tobe and Javier Fuentes-Mart\'{i}n 
for fruitful discussions and useful comments.
This work of K.~Y. was supported in part by the JSPS KAKENHI 18J01459.
The work of Y.~O. is supported by Grant-in-Aid for Scientific research from the Ministry of Education, Science, Sports, and Culture (MEXT), Japan, No. 17H05404.

\appendix

\section{Absorption of the phase of \mbox{\boldmath $C_{V_1}$}}
\label{phasedorp}

The global phase in \eqref{eq:Hamiltonian} is unphysical and hence can be reabsorbed.  
In our numerical study, we have taken $C_{V_1}$ to be real by absorbing its phase as illustrated below.

In the ${\rm S}_1$ LQ model, the relevant Wilson coefficients are $C_{V_1}, C_{S_2}$ and $C_T$.
A general formula for relevant observables is given
as a function of $( 1 + C_{V_1})$, $C_{S_2}$ and $C_{T}$  
  \begin{align}
  f\left(  C_{V_1},\,C_{S_2}, \,C_{T} \right)
  =&a_0 |1+C_{V_1}|^2 +a_1 |C_{S_2}|^2 +a_2 |C_{T}|^2 
  +a_3 \textrm{Re}[(1+C_{V_1})C_{S_2}^{\ast}]\nonumber  \\
  &+a_4\textrm{Re}[(1+C_{V_1})C_{T}^{\ast}]
  +a_5\textrm{Re}[C_{S_2} C_{T}^{\ast}],
  \end{align}
where $a_i$ are real constants.
Let us define $1 + C_{V_1} \equiv C_{0} e^{i \theta_0}$ where $C_0$ is a real dimensionless number, then one obtains 
  \begin{align}
  f\left(  C_{V_1},\,C_{S_2}, \,C_{T} \right)
  =&a_0 C_{0}^2 +a_1 |C_{S_2}|^2 +a_2 |C_{T}|^2 
  +a_3 C_0 \textrm{Re}[e^{i \theta_0} C_{S_2}^{\ast}]\nonumber  \\
  &+a_4 C_0 \textrm{Re}[e^{i \theta_0} C_{T}^{\ast}]
  +a_5\textrm{Re}[C_{S_2} C_{T}^{\ast}].
  \end{align}
The phase $ \theta_0$ can be absorbed by redefinitions of $C_{S_2}$ and $C_T$;
$C_{S_2} \to C_{S_2}^{\prime} = C_{S_2} e^{ - i \theta_0}$ and $C_{T} \to C_{T}^{\prime} = C_{T} e^{- i \theta_0}$. 
Besides, the LQ boundary condition $C_{S_2} (\mu_{\rm LQ})
= - 4  C_T (\mu_{\rm LQ})$ and the RG evolution
are compatible with the redefinitions.
Therefore, the independent parameters are only three: $C_0$, $|C_{S_2}^{\prime}(\mu_{\rm LQ})|$ and Arg$ \left[ C_{S_2}^{\prime}(\mu_{\rm LQ})\right]$.
This redefinition is also applicable for the case of the ${\rm U}_1$ LQ.


\end{document}